\documentclass[fleqn,usenatbib]{mnras}

\usepackage{newtxtext,newtxmath}

\usepackage[T1]{fontenc}
\usepackage{graphicx}	
\usepackage{amsmath}	
\usepackage{graphicx}
\usepackage{rotating}
\usepackage{capt-of}
\usepackage{adjustbox}
\usepackage{xspace}
\usepackage{comment} 

\usepackage{gensymb}
\newcommand{\fluxcgs}{erg~s$^{-1}$~cm$^{-2}$}
\newcommand{\source}{4U~1323--62\xspace}

\newcommand{\nustar}{\textit{NuSTAR}\xspace}
\newcommand{\xmm}{\textit{XMM-Newton}\xspace}

\newcommand{\src}{4U~1323--62\xspace}

\DeclareRobustCommand{\VAN}[3]{#2}
\let\VANthebibliography\thebibliography
\def\thebibliography{\DeclareRobustCommand{\VAN}[3]{##3}\VANthebibliography}

\usepackage{graphicx}	
\usepackage{amsmath}	
\usepackage[T1]{fontenc}
\usepackage{xspace}
\usepackage{multirow}
\usepackage{amsmath}
\usepackage{lipsum}
\usepackage{soul}
\usepackage[normalem]{ulem}
\usepackage{float}
\usepackage{comment}
\usepackage{hyperref}   

\title[\src]{Joint \xmm and \nustar Observations of Type I X-ray Bursts from the Dipping Low-Mass X-ray Binary \src}

\author[Boztepe, T. et al.]{Tu\u{g}ba Boztepe$^{1,3},$ 
Tolga G\"uver$^{2,3}$, 
Ersin G\"o\u{g}\"u\c{s}$^{4}$,
Diego Altamirano$^{5}$,
Julia Speicher$^{6}$,
Brian Luff$^{5}$,
\newauthor
David R. Ballantyne$^{7}$
\\
$^{1}$Istanbul University, Graduate School of Sciences, Department of Astronomy and Space Sciences, Beyazıt, 34452, \.{I}stanbul, T\"{u}rkiye \\
$^{2}$Istanbul University, Science Faculty, Department of Astronomy and Space Sciences, Beyaz{\i}t, 34452, \.{I}stanbul, T\"{u}rkiye \\
$^{3}$Istanbul University Observatory Research and Application Center, Istanbul University 34452, \.{I}stanbul, T\"{u}rkiye \\
$^{4}${Faculty of Engineering and Natural Sciences, Sabanc\i~University, Orhanl\i-Tuzla 34956, \.Istanbul, T\"urkiye}\\
$^{5}$School of Physics and Astronomy, University of Southampton, Southampton SO17 1BJ, UK \\
$^{6}$Anton Pannekoek Institute for Astronomy, University of Amsterdam, Postbus 94249, 1090 GE Amsterdam, The Netherlands \\
$^{7}$Center for Relativistic Astrophysics, School of Physics, Georgia Institute of Technology, 837 State Street, Atlanta, GA 30332-0430, USA}
\date{Accepted XXX. Received YYY; in original form ZZZ}

\pubyear{\the\year{}}

\begin{document}
\label{firstpage}
\pagerange{\pageref{firstpage}--\pageref{lastpage}}
\maketitle

\begin{abstract}
In this study, we report partly simultaneous \xmm and \nustar observations of the bursting, dipping low mass X-ray binary, \src obtained in 2024. \src is one of the well-known persistent bursters, with bursts occurring roughly every three hours. It is also one of the few sources for which the orbital period is known, and shows dips in X-rays. In this paper, we report the detection of 12 unique bursts with \xmm and \nustar, 6 of them observed jointly. We detected two double burst events, one with the \nustar and another one observed with both missions. During our observations we detected 10 X-ray dips with a periodicity of 2.942 hours, in line with previous measurements. 
We also present the results of the time resolved X-ray spectral analysis of the bursts and show the limits on the cooling of the corona heated by the burst emission. We also found a $0.898 \pm 0.017$~Hz quasi-periodic oscillation~(QPO) during the non-bursting and non-dipping times confirming previous detections.
\end{abstract}



\section{Introduction}

Low mass X-ray Binaries~(LMXBs) are systems consisting of a compact object - either a black hole or a neutron star - that accretes material from a low-mass companion star ($M \lesssim 1 M_{\odot}$) via Roche-lobe overflow. The gravitational potential energy released during accretion powers the X-ray emission observed from these systems. \src is one of the early members of the bursting low mass X-ray binaries group. It was first discovered by Uhuru and Ariel V \citep{Forman_1978,Warwick_1981}. Bursts and periodic intensity dips in its X-ray lightcurves were first detected by EXOSAT \citep{vander_7600607,Parmar_7600607}. Dips from the source are observed to last for about 30\% of the orbital cycle \citep{Boirin_7600607} and have an energy dependence. The dips can be as deep as 55\% in the lower energy bands~(0.5--2.5 keV) and reach to about 36\% in higher energy bands~(2.5--10 keV). Similarly, lightcurves from SUZAKU PIN show dips up to 20~keV with a depth of about 10\% $\pm$ 6\% \citep{2009A&A...500..873B}. 
 
In addition to the bursting group, the presence of dips in the X-ray light curve of \source also grants it membership of the group of LMXBs known as dippers. There are currently 29 such sources known, whose dipping phenomenology is variable. The widely accepted explanation of the absorption dips in the light curve of a dipper was advanced by \cite{1987A&A...178..137F}, that a bulge may develop at the point where the accretion flow impacts the edge of the disc. There is a limit to how far such a bulge could extend, azimuthally, and that constrains the inclination of the orbital plane to our line of sight to the system. Unlike the majority of dippers that dip sporadically and aperiodically, \source is among the very best-behaved, such that its inclination may safely be constrained to be between 60\degree -- 80\degree~\citep{1987A&A...178..137F,Parmar_7600607}. The orbital period of the source is constrained to be around P = 2.94~hours with a secular decreasing trend likely due to gravitational radiation \citep{Gambino_7600607}. The mass of the companion is determined as 0.28~M$_{\odot}$ and the distance of the source is estimated as 4.2~kpc \citep[see, e.g.,][]{Gambino_7600607, Zolotukhin_7600607} using Galactic reddening maps and the extinction towards \source.

Type I X-ray bursts~(hereafter X-ray bursts) are thermonuclear explosions occurring in the surface layers of weakly magnetized neutron stars that are accreting mass from a companion star \citep[see e.g.,][]{1993SSRv...62..223L,2008ApJS..179..360G,2020ApJS..249...32G}. The accretion process typically involves matter (hydrogen and/or helium) steadily transferred from the companion star, often via an accretion disk, onto the NS surface.
As the matter accumulates, it reaches a critical density and temperature, triggering an unstable thermonuclear flash of H/He. The resulting burst produces a characteristic light curve, a fast rise time (typically $\sim$1-10 s) followed by a slower decay phase lasting tens to hundreds of seconds, during which the neutron star surface cools. 

The intense burst of X-rays strongly irradiates the surrounding environment, including the accretion disk and its hot corona. During an X-ray burst, the soft X-ray photons from the neutron star surface can rapidly cool the hot coronal electrons via Compton scattering, leading to a temporary reduction of the hard X-ray emission from the corona \citep{2003A&A...399.1151M,chen2012,degenaar2018,buisson2020,koljonen2023}.
This process, often termed coronal cooling, provides a powerful diagnostic to probe the physical properties and geometry of the corona itself \citep{2000ApJ...537L.107M,2002MNRAS.335..465M,2003A&A...399.1151M,2020MNRAS.499.4479S}.

A persistent $\sim$1 Hz QPO was first detected in \source by \cite{1999ApJ...511L..41J}, but its physical origin remains unclear. \cite{2000ApJ...542L.111T} proposed that this QPO arises from global normal-mode oscillations of the accretion disc driven by the gravitational interaction with the neutron star. \cite{2010MNRAS.407.1895B} later identified two QPO components at 1.4 and 2.8~Hz through frequency-resolved spectroscopy, linking them to a reflection region in the inner accretion disc. \cite{Bhulla_2020} subsequently confirmed the presence of this feature at $\sim$0.97 Hz using AstroSat/LAXPC, with an rms amplitude of 6.7\% and a quality factor of $\sim$18, consistent with previous detections.

Despite being observed with multiple missions over several decades, the simultaneous broadband coverage of \xmm and \nustar provides a unique opportunity to study the persistent emission, dipping behavior, and thermonuclear X-ray bursts of \source in unprecedented detail. In particular, the hard X-ray coverage of \nustar allows us to search for evidence of coronal cooling during the bursts, which has not been possible with previous observations of this source.

This paper is organized as follows. \hyperref[sec:bursts]{Section~\ref*{sec:observation_datareduction}} outlines the observation and data reduction. \hyperref[sec:bursts]{Section~\ref*{sec:persistent}} details the spectral analysis of the persistent emission and the search for quasi-periodic oscillations (QPOs). In \hyperref[sec:bursts]{Section~\ref*{sec:bursts}}, we present the detection and time-resolved spectral analysis of the thermonuclear X-ray bursts, including a search for evidence of coronal cooling. \hyperref[sec:bursts]{Section~\ref*{sec:dips}} describes the X-ray dips detected during the observations and their spectral properties. Finally, in \hyperref[sec:bursts]{Section~\ref*{sec:conclusion}} we discuss our findings in the context of previous observation of \source.

\section{Observations and Data Reduction}
\label{sec:observation_datareduction}

We observed \source on 7 August 2024, quasi-simultaneously with \nustar and \xmm for approximately 120~ks. In \autoref{tab:observations}, we provide the details of individual observations. To detect all the temporal variations in the data and match the datasets, we applied barycentric correction to the event files of both \xmm and \nustar using the JPL-DE440 solar system ephemeris and source coordinates as RA=201.5592~(J2000) and DEC=-62.1443~(J2000).

\autoref{fig:all_lc} shows the combined lightcurve of both observations. The \xmm~(blue) and unfiltered lightcurve of \nustar~(orange), which also shows the missed bursts in clean event files, are shown for the 0.5--10 keV and 3--30 keV ranges, respectively. 

In  \hyperref[sec:xmmanalysis]{Section~\ref*{sec:xmmanalysis}} and \hyperref[sec:nuanalysis]{Section~\ref*{sec:nuanalysis}}, we describe the data reduction and calibration procedures.

\subsection{XMM-Newton}
\label{sec:xmmanalysis}

The \xmm data for this source consist of a single observation started on 2024 August 7 (OBSID 0935800201). All EPIC instruments were active (pn, MOS1, MOS2) along with the RGS during the observation. Both pn and MOS2 detectors were operated in timing mode, while MOS1 was in small window mode during the observation. From the EPIC-pn data set, we extracted source counts which fell within the RAWX detector columns 29–47 (inclusive), corresponding to a width of 19 pixels (or an effective extraction width of $\sim$ 77.9 arcsec).
We performed data reduction using  the Science Analysis System (SAS) version 22.1.0, following the standard analysis threads. We used the current \xmm calibration files as of June 2025. 
The EPIC data were screened for background flares and photon pile-up. High background intervals in the MOS1 data ($\sim 17\%$) were excluded using standard GTI filtering, while MOS2 and PN showed negligible flaring activity. No significant pile-up was detected in any of the EPIC instruments. The EPIC-pn and MOS event files are generated using \texttt{epproc} and \texttt{emproc}, respectively. Due to the low count rate of the source, we here focus mainly on the EPIC-pn data. The \texttt{evselect} routine is used to generate the final product for the source and background light curves and spectra. For the source, we selected all the events with PATTERN smaller than 4 and in between 29th and 47th columns in RAWX coordinate format. For the background we selected similar events from a region away from the source. \texttt{rmfgen} and \texttt{arfgen} are used to generate spectral response files for each spectrum. We extracted source products in energy range 0.5--10.0~keV. 

\subsection{NuSTAR}
\label{sec:nuanalysis}
The \nustar observation was performed on 2024 August 7 (OBSID 31001021002) in the Science Mode. The \nustar Observatory consists of two co-aligned, identical telescopes that focus hard X-rays in the 3--79~keV energy band onto two independent focal plane modules, FPMA and FPMB \citep{2013ApJ...770..103H}. The total ontime of the \nustar observation we have is 154.79~ks (see \autoref{tab:observations}) corresponding to a clean exposure time of 90.1~ks. For the analysis of the \nustar data, we utilized the \texttt{nupipeline} and \texttt{nuproducts} tools within HEASOFT v6.35, with the calibration files available as of 2024 August 12. We extracted source and background spectra and the corresponding redistribution matrices and ancillary response files using the nuproducts pipeline.\\
We used circular regions to select source events from both FPMA and FPMB detectors. The source counts were extracted from a circular region with a radius of $\sim 80''$ centered on the source position (RA: 201.6548653, DEC: -62.1350533). For background subtraction, another circular region with a radius of $\sim104.4''$ was selected from a source-free area on the same detector (RA:201.6991971 DEC:-62.2442650).

Similar to the \xmm data, we note that there are no instrumental effects in the NuSTAR observations that need to be taken into account. We also utilized the unfiltered event files in the initial detection of the thermonuclear X-ray bursts, which allowed us to detect two more burst, shown in \autoref{tab:nustar_burst_values}.

\begin{figure*}
\centering
\includegraphics[scale=0.55]{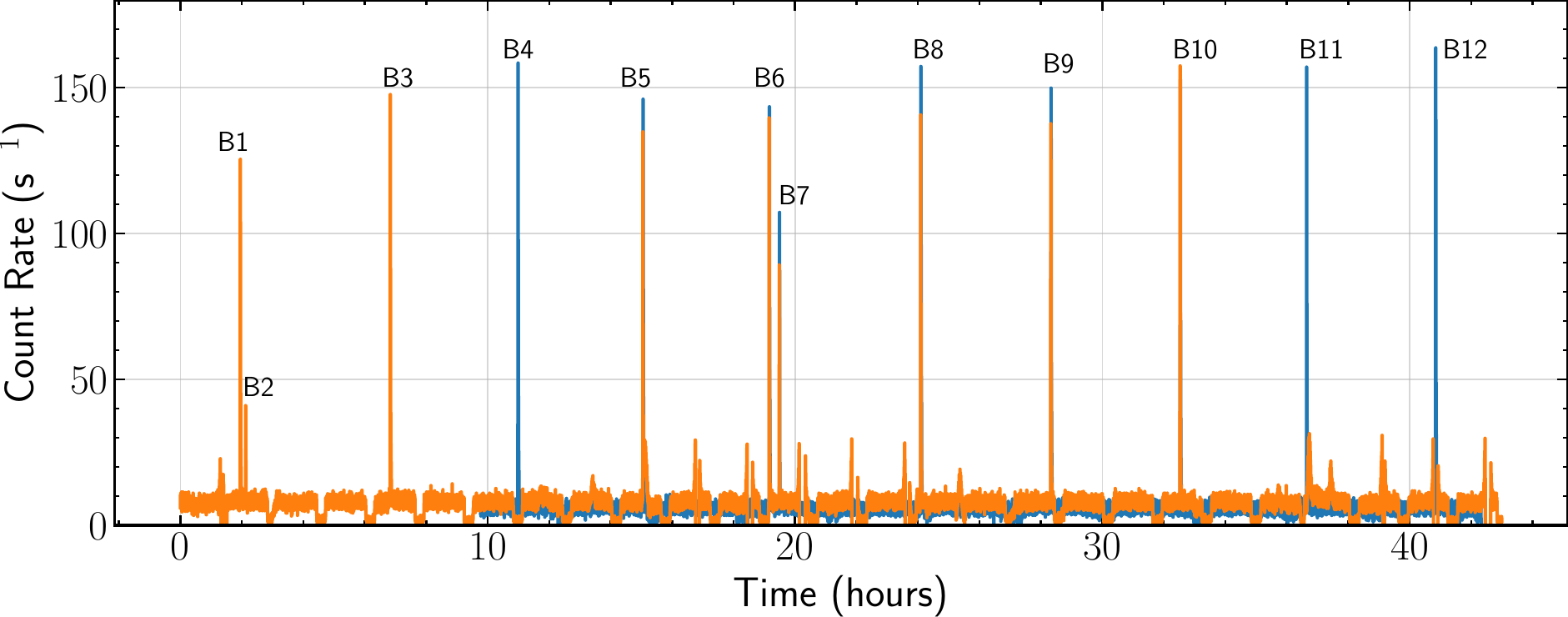}
\caption{0.5--10~keV and 3--30~keV lightcurves of \source obtained with \xmm (blue) and \nustar (orange), respectively. Time is shown in units of hours after the start of the \nustar observation. To be able to show all the bursts observed with \nustar, we here show the lightcurve extracted from the unfiltered event file. Therefore intervals with increased background are also visible, real thermonuclear bursts have at least 29 counts/s at the peak.}
\label{fig:all_lc}
\end{figure*}

\begin{table*}
\caption{Details of the observations of \src using \nustar and \xmm. For \xmm, the parameters correspond specifically to the EPIC-pn detector.}
\begin{tabular}{cccccc}
\hline 
Instrument& Start Time&Stop Time& Net Exp.&Obs. ID\\
&(MJD)&(MJD)&(ks)&\\
\hline
\nustar&60529.29249634&60531.08410514&90&31001021002&\\
\xmm& 60529.69959637&60531.06053387&135.5&0935800201&\\ 
\hline
\end{tabular}
\label{tab:observations}
\end{table*}

\section{Results}

Having the data ready, below we present our methods and results from different analyses related to various features observed from \src. We start with spectral and temporal analysis of the persistent non-bursting, non-dipping states of the source. We then continue with identification and time resolved spectra analysis of the thermonuclear X-ray bursts. Having broadband simultaneous data, we also search for any evidence for coronal cooling during the bursts. Finally we focus on the X-ray dips detected in the \xmm data.

\subsection{X-ray Spectral and Temporal Properties of the Persistent emission}
\label{sec:persistent}
\subsubsection{X-ray Spectroscopy}

As can be seen in \autoref{fig:all_lc} and \autoref{fig:dips} in total we detected 12 unique X-ray bursts and 10 X-ray dips within the observations (see \hyperref[sec:bursts]{Section~\ref*{sec:bursts}} and \hyperref[sec:dips]{Section~\ref*{sec:dips}}). In order to determine X-ray spectral characteristics of the source during the non-bursting, non-dipping times we created an X-ray spectrum using the data where we excluded all the burst and dip times given in \autoref{tab:xmm_burst_values}, \autoref{tab:nustar_burst_values} and \autoref{tab:dip_times}, respectively. The resulting data had 94.6 ks and 75.6~ks of exposure time for \xmm and \nustar, respectively. We extracted X-ray spectra and grouped to have at least 50 counts in each channel. 

Using \texttt{xspec} \citep{1996ASPC..101...17A}, we fitted the data with an absorbed \texttt{comptt} model. For the interstellar absorption, we used the \texttt{tbabs} model with ISM abundances \cite{wilms2000absorption-574}. Following the inter-calibration notes\footnote{\url{https://xmmweb.esac.esa.int/docs/documents/CAL-TN-0230-1-3.pdf}\label{calnote}} between the \nustar and \xmm EPIC pn camera, we added a constant flux offset parameter, keeping the \nustar values fixed to 1 and allowing EPIC pn parameter to vary. 

\begin{figure*}
    \centering
    \includegraphics[scale=0.45]{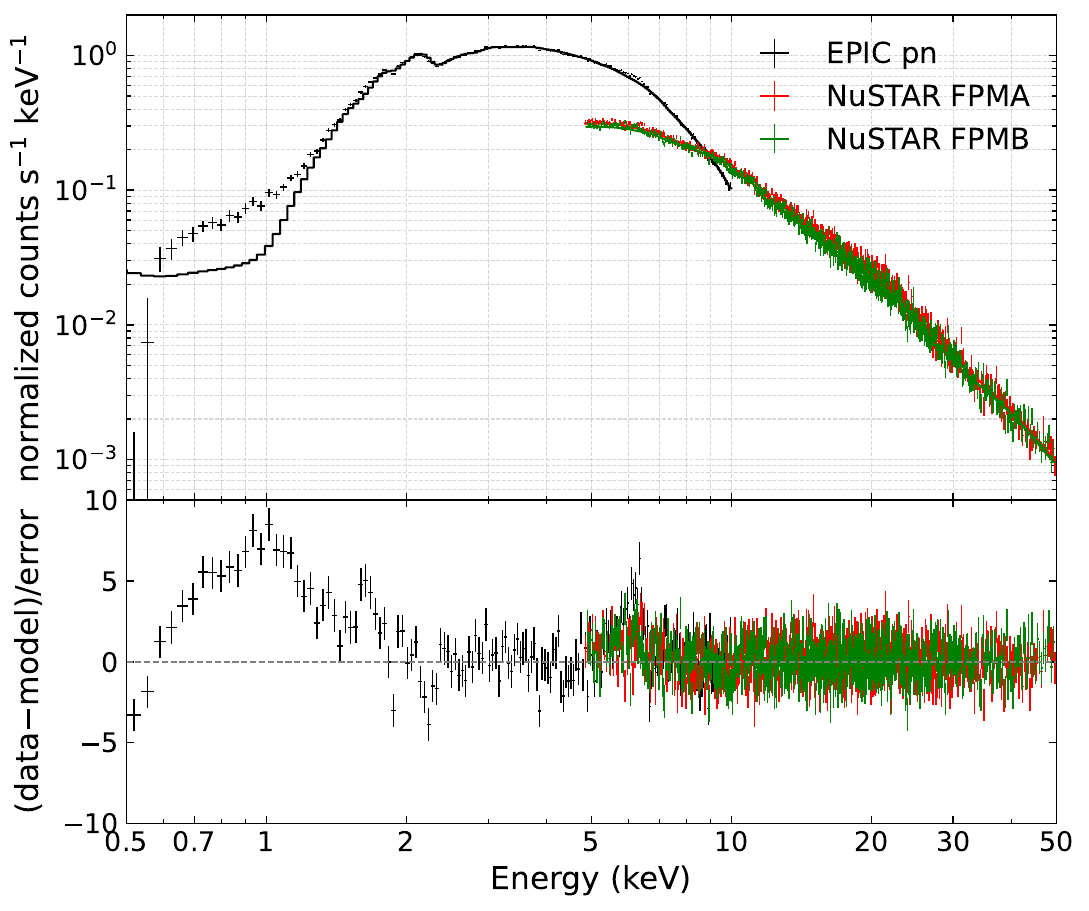}
    \includegraphics[scale=0.45]{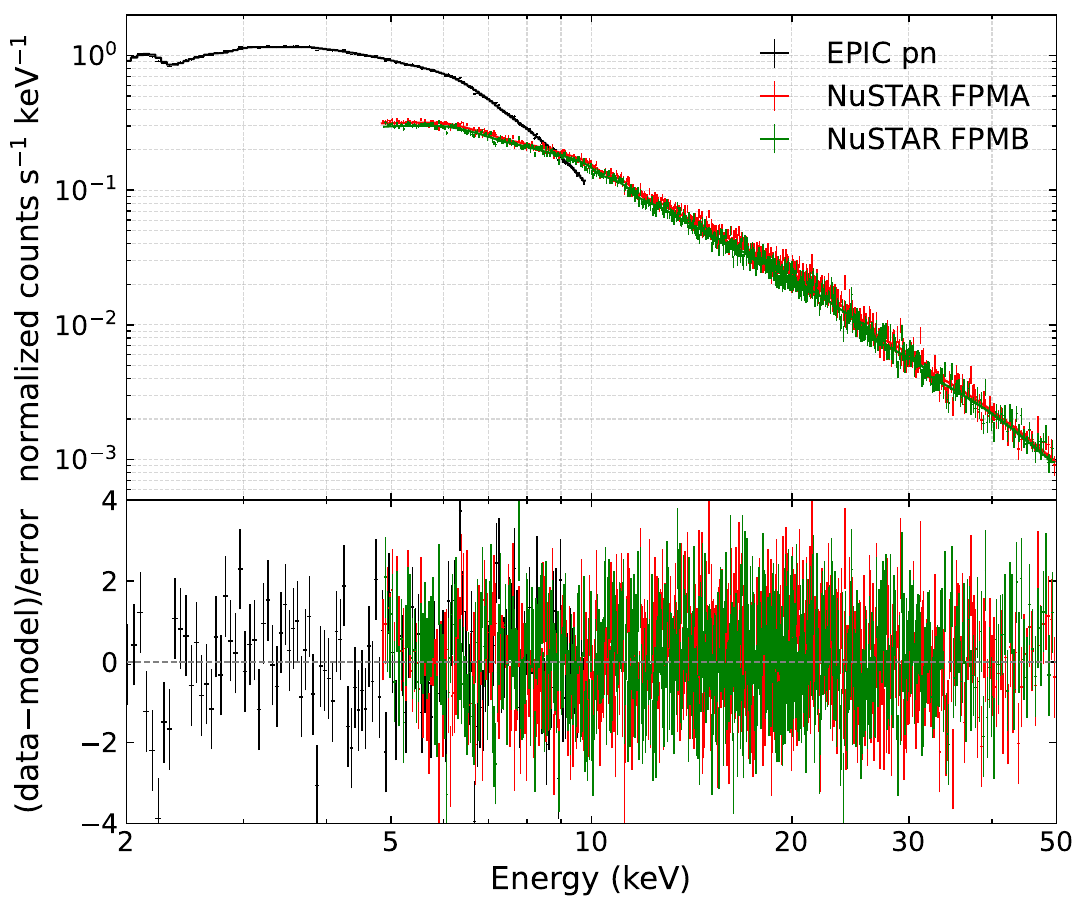}
\caption{{\it Left Panel:} X-ray spectra of \source extracted from the non-dipping and non-bursting times obtained with \xmm (black) and \nustar (red and green for FPMA and FPMB). Similar to \protect\cite{Boirin_7600607} we also see excess emission at around 1~keV region. Additionally an emission line at around the 6 keV region is also visible. {\it Right Panel:} The best fit model with the \xmm and \nustar data with residuals shown. The best fit model now has a gaussian line. In each panel, lower panels shows the residuals from the best-fit model.}
\label{fig:pers_spec}
\end{figure*}

Initially we fitted the 0.5--10.0~keV range in the \xmm and 5.0--50.0~keV range in the \nustar data using the above model (\texttt{constant * tbabs * comptt}). As shown in \autoref{fig:pers_spec}, we see a significant excess at around 1~keV range in the EPIC pn data. This excess was also seen by \cite{Boirin_7600607}, in the EPIC pn data but not in the RGS data. Following \cite{Boirin_7600607}, we also wanted to check the RGS data, however the fact that during our observations \src was  significantly dimmer, RGS data did not provide any conclusive information in that manner. We therefore, ignored the EPIC pn data below 2~keV, following \cite{Boirin_7600607}. In addition to this, we also detect an emission line feature at around 6~keV. We modeled this component with a Gaussian emission line. This way we can obtain an acceptable fit in the 2.0 -- 50.0~keV range using \xmm EPIC pn and \nustar FPMA and FPMB data, with a reduced $\chi^2_{\nu}$ = 1.13 for 1322 degrees of freedom. Adding a Gaussian line centered at $6.01\pm0.05$~keV improves the fit quality by $\Delta\chi^2$ = 131.12 for 3 degrees of freedom, confirming its statistical necessity at >5$\sigma$ confidence level. The X-ray spectra and the best fit models are shown in \autoref{fig:pers_spec}. All the parameter uncertainties are  calculated for 1$\sigma$ range   using Xspec \emph{error} command.

\begin{figure}
    \centering
    \includegraphics[width=1.0\linewidth]{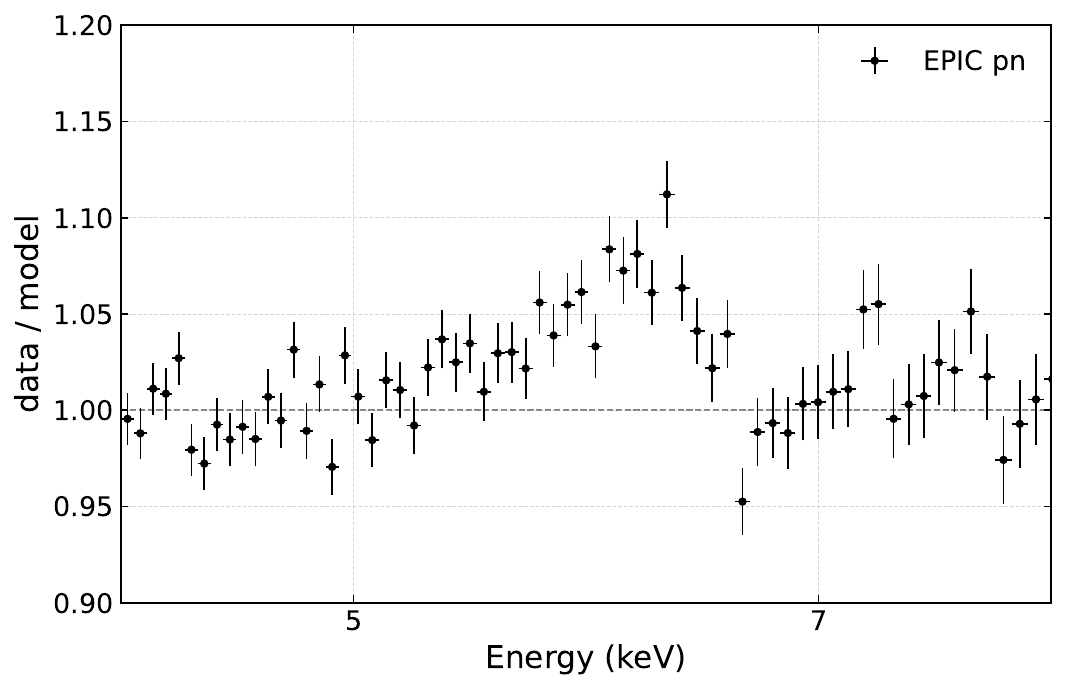}
    \caption{Ratio of the best fit model (\texttt{tbabs*comptt}) to the data emphasizing the shape of the Gaussian line feature.}
    \label{fig:gaussline}
\end{figure}

Our analysis reveals that the absorbed X-ray flux of \src in the 2.0--10.0~keV and 5.0--50.0~keV range is (6.57$\pm0.02)\times10^{-11}$~\fluxcgs and (2.23$\pm0.01)\times10^{-10}$~\fluxcgs, respectively. The resulting best fit parameters are as follows; for the flux offset between EPIC pn and FPMA and FPMB we found that the EPIC pn measures fluxes about 20.0$\pm$0.2\% lower compared to \nustar. This value is in very good agreement with the offset found in inter-calibration studies$^{\ref{calnote}}$. We inferred that the hydrogen column density in the line of sight towards \source is (3.36$\pm0.09)\times10^{22}~\rm{cm}^{-2}$. As for the \texttt{comptt} model, we found that seed photons had a temperature of 0.85$\pm$0.01~keV, while the plasma temperature was found to be 13.83$\pm$0.48~keV, and the optical depth to inverse Compton scattering $\tau$=2.90$\pm$0.06. As for the Gaussian line, we find that the line energy is at 6.02$\pm$0.05~keV, with a width of $\sigma$=0.42$\pm$0.07~keV. We note that the line feature we detect is highly asymmetric as shown in \autoref{fig:gaussline}, we tried to fit the feature with models like \texttt{laor} \citep{1991ApJ...376...90L} or \texttt{diskline} \citep{1989MNRAS.238..729F} but obtained similar results. Finally, the unabsorbed total flux of the source we obtained is (9.45$\pm0.03)\times10^{-11}$~\fluxcgs in the 2.0--10.0~keV range, while in the 5.0--50.0~keV range it is (2.26$\pm0.01)\times10^{-10}$~\fluxcgs.

\subsubsection{Quasi-periodic oscillations (QPOs) in the persistent emission}
\label{sec:qpo}

\src is one of the few sources that exhibit both dipping and bursting behaviors; therefore, its persistent emission provides a unique opportunity to search for different types of quasi-periodic oscillations~(QPOs). In the following, we investigate the data for two specific features. The firs one is the $\sim$1~Hz QPO, which was detected in a number of observations before from this source. In addition to the 1~Hz feature, we also searched for millihertz quasi-periodic oscillations~(mHz QPOs), which are typically associated with X-ray bursters.

\begin{figure}
    \centering
    \includegraphics[width=0.8\linewidth]{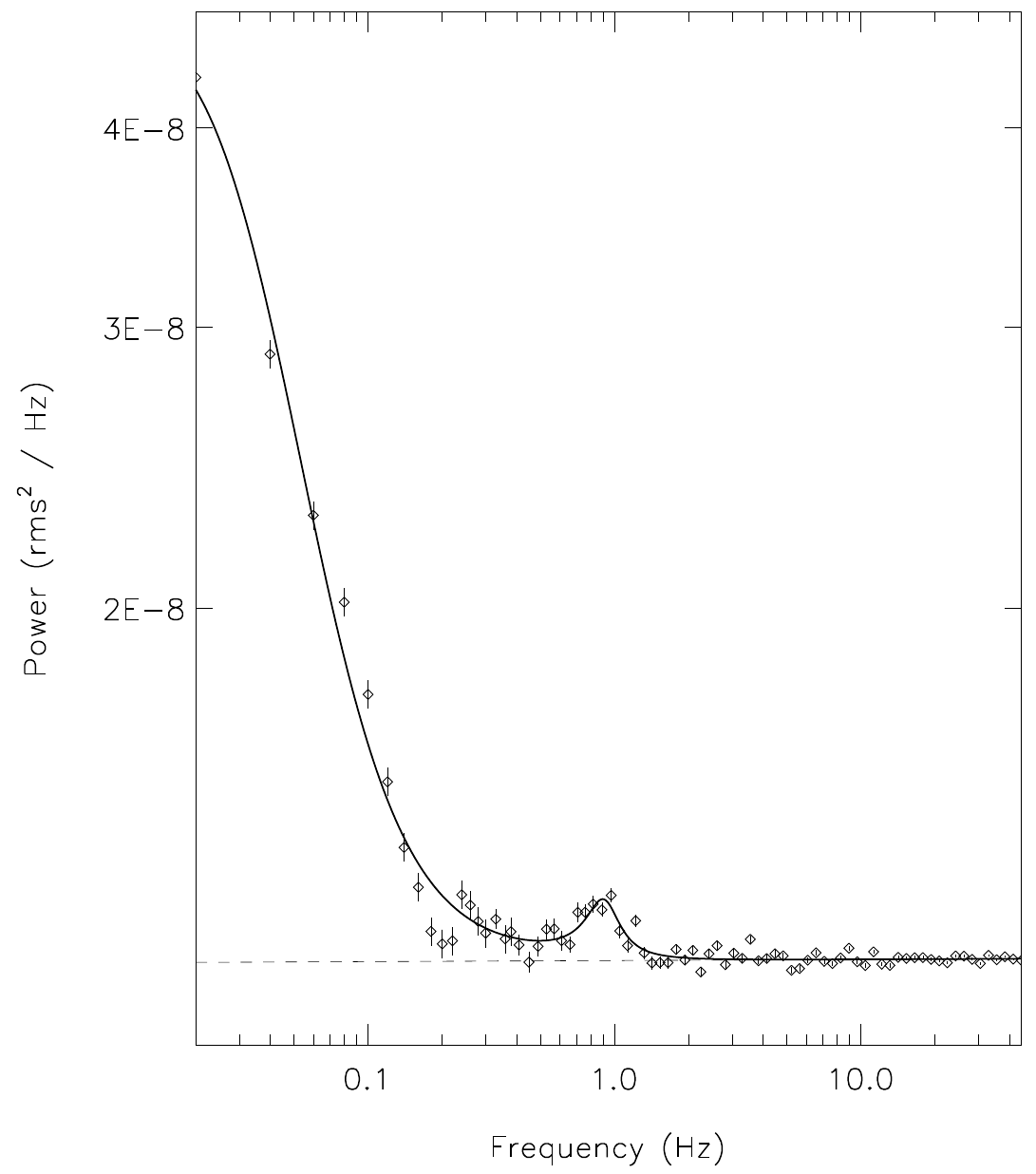}
    \caption{The averaged power spectrum of the persistent emission of \src. 
    }
    \label{fig:1hz}
\end{figure}


To search for this QPO in the persistent emission, we extracted a light curve from the EPIC-pn data in the 0.5--10.0~keV energy band with a time resolution of 0.01~s. Removing the dips and bursts from the data we created power spectra utilizing Fast Fourier Transforms of 50~s segments, yielding a total of 2343 segments. The resulting average power spectrum was rebinned logarithmically to 30 bins per frequency decade and fitted with a combination of a power-law and a Lorentzian. As shown in \autoref{fig:1hz}, we detected a broad QPO centered at $\nu = 0.898 \pm 0.017$~Hz with a full width at half maximum (FWHM) of $0.355 \pm 0.052$~Hz. The integrated power of the Lorentzian component yields a fractional rms amplitude of  $ \sim$2.4\%, which is consistent with previous studies of the source (see \cite{1999ApJ...511L..41J,Bhulla_2020}.

Within the persistent times of the data we also searched for oscillations at mHz frequencies by computing the power density spectra (PDS) from the EPIC pn data. Because these oscillations are mostly observed in the soft X-ray bands \citep[see, e.g.,][]{2008ApJ...673L..35A,2019MNRAS.486L..74M}, we created light curves in the 0.5--5.0 and 2.0--5.0~keV energy bands at time resolutions of 10, 12 and 16~s. We divided the data into segments where no dip or burst is observed and last at least 4000~s.  

Following \citet{2005A&A...431..391V}, we fitted the log--log PDS with a broken power-law noise model and derived 99\% confidence levels from the $\chi^{2}$ distribution, accounting for the number of independent frequency bins searched. We searched 2--15 mHz frequency range, motivated by the 8.1 mHz QPO previously detected in \src with \textit{RXTE} \citep{2011ATel.3258....1S} and the typical mHz QPO range of $\sim$2--14 mHz observed in other neutron-star low-mass X-ray binaries \citep{2008ApJ...673L..35A,2019MNRAS.486L..74M,2021MNRAS.500...34T}. 

Our results showed that within 99\% confidence there is no significant mHz oscillation component detected in our data. The analysis provides an upper limit of 2--4\% on the fractional rms amplitude in the 2–15 mHz range, likely due to the relatively low count rate of the \source during our observation.

\subsection{Thermonuclear X-ray Bursts}
\label{sec:bursts}

During both observations, a total of  18 bursts~(6 of which are simultaneous) are already visible by eye~(see \autoref{fig:all_lc}). For the determination of the start times of Type-I X-ray bursts we applied the Bayesian Blocks algorithm~\citep[see, e.g.,][]{1998ApJ...504..405S,2012ascl.soft09001S}. For \nustar, we utilized the combined data from both modules (FPMA and FPMB) in the 3–30 keV energy range, while for \xmm, the search was focused on the 1-10 keV band. We configured the algorithm with a geometric prior to prevent over-segmentation and a significance factor of 3. The start of a  burst was determined if the count rate showed a significant increase over the preceding block and exceeded a minimum threshold of 15-20 cts/s. We determined the burst start times in Modified Julian Date~(MJD) with a 1.0 second time resolution. 

This way, we were able to detect 9 thermonuclear X-ray bursts in the \xmm data while 7 events in the clean \nustar data. In addition to these checks we also searched for the unfiltered \nustar event files for additional bursts and identified two more bursts all of which are also simultaneously observed with \xmm. As shown in \autoref{tab:xmm_burst_values} and \autoref{tab:nustar_burst_values} in total we are able to detect 6 bursts simultaneously with both satellites. Two of these simultaneously detected bursts are only seen in unfiltered \nustar data. Because \texttt{nupipeline} and \texttt{nuproducts} only creates and works on cleaned event files we only note these bursts here, but do not perform further spectral analysis. We assign unique burst ID numbers for each event starting from the first burst detection with \nustar within our observing campaign. In \autoref{fig:all_lc} we show these burst ID next to each thermonuclear burst event.

Once the start time of each burst is determined, we generated light curves with 2 second time bins and for each burst, we identified the peak count rate. The rise time was calculated as the time reached from the beginning of the burst to the point where the count rate first reached 90\%  of the peak value.  Finally, to characterize the burst decay, we fit an exponential function to the post peak light curve, yielding the e-folding decay time and its uncertainty. All measured values with \xmm and \nustar are given in \autoref{tab:xmm_burst_values} and \autoref{tab:nustar_burst_values}. \autoref{fig:xmmburst_lc} and \autoref{fig:nustarburst_lc} show the lightcurves of detected bursts by \xmm in the 0.5--10.0~keV and \nustar in the 3--30~keV band, respectively.

In general, bursts show similar profiles, with peaks typically occurring with comparable timescales. The observations reveal two double bursts shown in \autoref{fig:nustarburst_lc}. The left panel in \autoref{fig:nustarburst_lc} displays a double-burst event detected only by \nustar (bursts \#1 and \#2), before the start of the \xmm exposure. In this instance, the two peaks are separated by $\sim$628 seconds. While the primary burst reaches a peak intensity of 136 cts/s, the secondary ignition is significantly weaker (52 cts/s) in the 3--30~keV band. The ratio of the peak count rates show that the peak intensity of the second burst is only 38\% of the initial burst. The right panel shows a second simultaneous double-burst event captured by both \xmm and \nustar (one burst is in unfiltered \nustar event file). In this second case the separation between the start times is 1193 seconds and this time the peak count rate of the second burst is 70\% of the first peak in the \nustar band while the change in the peak count rate is similar in the \xmm band with a ratio of about \%68. 

\begin{table*}
\centering
\caption{Overview of X-ray bursts from \source observed with \xmm in the 0.5--10~keV range.}
\begin{tabular}{ccccccc}
\hline 
Burst&Start date&Rise Time&Peak Duration&Peak Count Rate& Pre-burst Count Rate & e-folding Time \\
ID$^{a}$&MJD& (s)&(s)&rate (c~s$^{-1}$)&rate (c~s$^{-1}$)&(s)\\
\hline

4 & 60529.750354& 7.5 & 1.0 & $199.0 \pm 14.1$ & $6.1 \pm 0.5$ & $14.9 \pm 0.5$ \\
5 & 60529.919834&5.5 & 7.0 & $171.0 \pm 13.1$ & $5.7 \pm 0.4$ & $15.6 \pm 0.7$ \\
6 & 60530.091026&5.5 & 3.0 & $180.0 \pm 13.4$ & $6.2 \pm 0.4$ & $17.0 \pm 0.6$ \\
7 & 60530.104834&1.5 & 5.0 & $121.0 \pm 11.0$ & $6.0 \pm 0.4$ & $13.0 \pm 0.3$ \\
8 & 60530.296489&7.5 & 6.0 & $168.0 \pm 13.0$ & $6.0 \pm 0.4$ & $23.8 \pm 0.7$ \\
9 & 60530.472947&6.5 & 5.0 & $167.0 \pm 12.9$ & $5.7 \pm 0.3$ & $21.6 \pm 0.6$ \\
10 & 60530.648236&3.5 & 8.0 & $177.0 \pm 13.3$ & $5.3 \pm 0.3$ & $23.9 \pm 0.6$ \\
11 & 60530.819405&5.5 & 5.0 & $171 \pm 13.15$ & $6.1 \pm 0.4$ & $18.6 \pm 0.5$ \\
12 & 60530.994498&4.5 & 3.0 & $182.0 \pm 13.5$ & $5.8 \pm 0.4$ & $19.0 \pm 0.5$ \\
\hline
\end{tabular}
\label{tab:xmm_burst_values}\\
\footnotesize{$^a$ This is the unique burst ID based on the numbering of the events since the  start of our observing campaign.}\\
\end{table*}

\begin{table*}
\centering
\caption{Overview of X-ray bursts from \source observed with NuSTAR. The parameters are obtained from 3--30~keV lightcurves with 0.5~s binning.}
\begin{tabular}{ccccccc}
\hline 
Burst & Start date& Rise Time& Peak Duration& Peak Count& Pre-burst Count& e-folding Time \\
ID$^{a}$ & MJD &(s) &(s) &rate (c~s$^{-1}$) &rate (c~s$^{-1}$) & (s) \\
\hline
1 & 60529.374111 &4.5 & 4.0 & $136.0 \pm 11.7$ & $2.2 \pm 0.3$ & 27.1$\pm$0.1 \\
2 & 60529.381380&4.5 & 0.0 & $52.0 \pm 7.2$ & $3.3 \pm 0.3$ & 17.2$\pm$0.3 \\
3 &60529.577421&3.5 & 2.0 & $165.0 \pm 12.8$ & $4.2 \pm 0.4$ & 27.8$\pm$0.1\\
5 &60529.919840&5.5 & 4.0 & $141.0 \pm 11.9$ & $3.1 \pm 0.3$ & 21.4$\pm$0.1 \\
6$^*$ &60530.089931 &4.5 & 7.0 & $147.0 \pm 12.1$ & $2.3 \pm 0.3$ & 24.2$\pm$0.1 \\
7 & 60530.104829&1.5 & 4.0 & $103.0 \pm 10.2$ & $2.9 \pm 0.3$ & 15.1$\pm$0.1 \\
8$^*$ &60530.295417  &5.5 & 4.0 & $160.0 \pm 12.7$ & $2.4 \pm 0.4$ & 27.1$\pm$0.1 \\
9 & 60530.472942&3.5 & 6.0 & $150.0 \pm 12.2$ & $1.9 \pm 0.3$ & 25.9$\pm$0.1 \\
10 & 60530.648232&4.5 & 3.0 & $164.0 \pm 12.8$ & $1.7 \pm 0.3$ & 23.1$\pm$0.1 \\
\hline
\end{tabular}
\label{tab:nustar_burst_values}\\
\footnotesize{$^a$ This is the unique burst ID based on the numbering of the events since the  start of our observing campaign.\\}
{$^*$ These bursts are only seen in unfiltered event files and the start times are not barycentered.}\\
\end{table*}

\begin{figure*}
\begin{centering} 
    \includegraphics[scale=0.28]{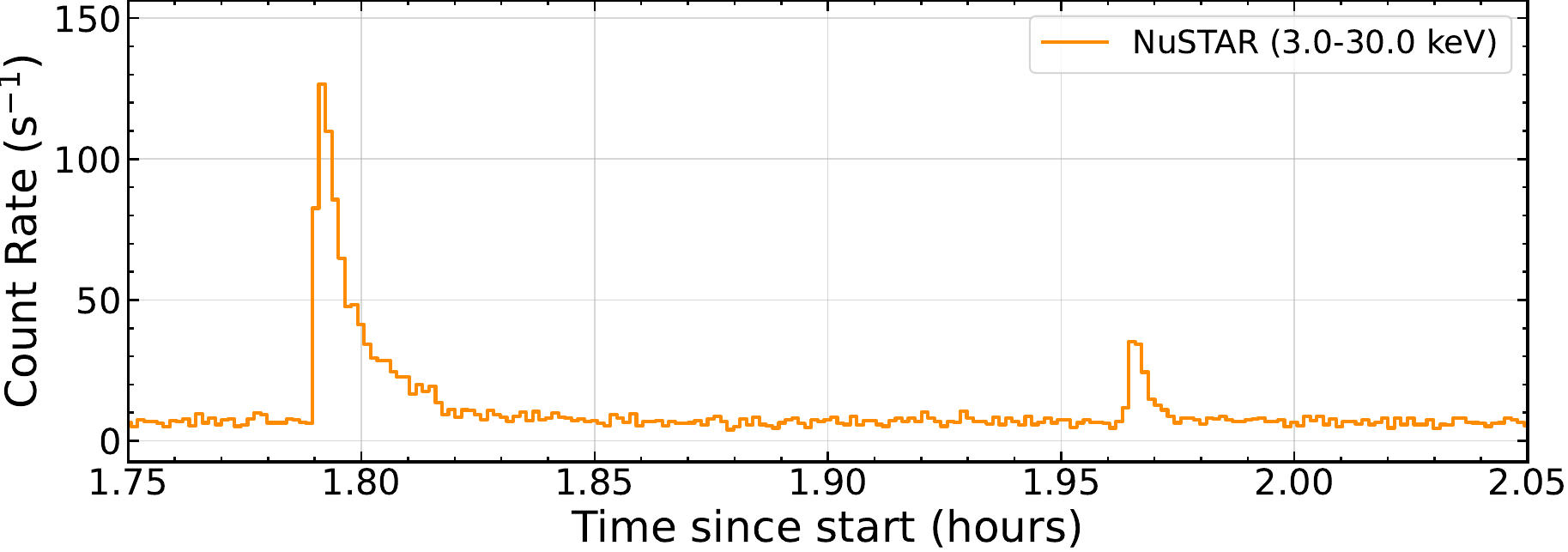}
\includegraphics[scale=0.28]{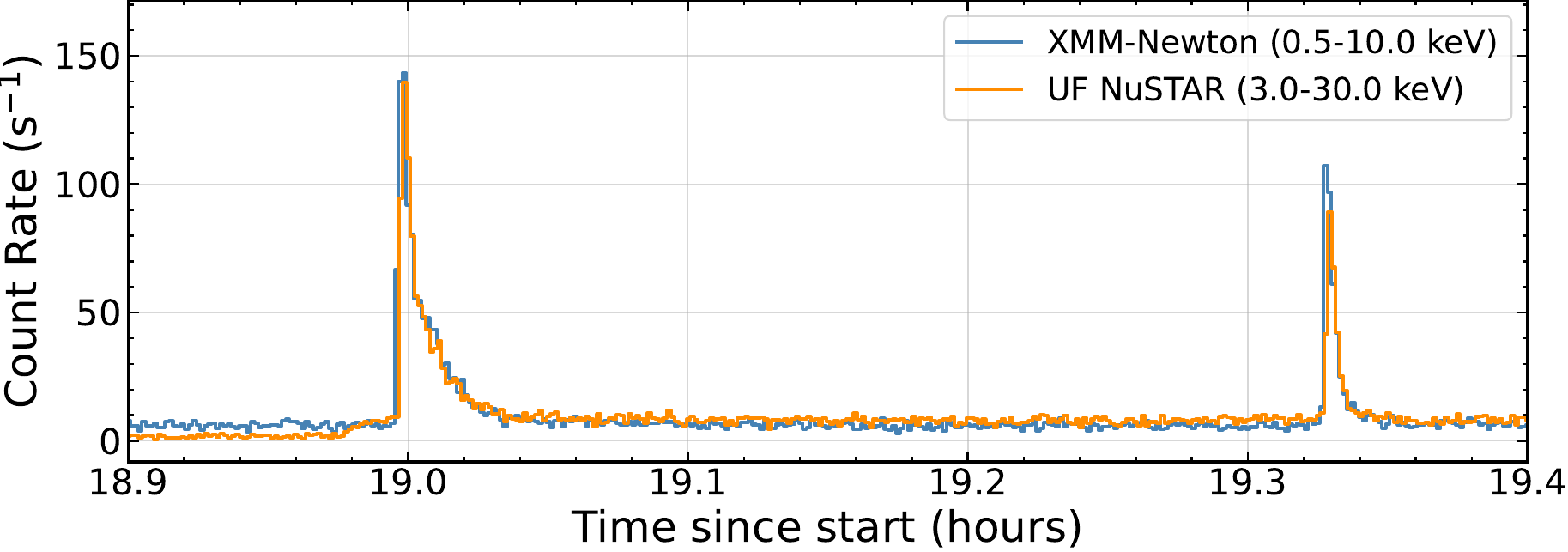}
    \caption{Double bursts observed with \nustar and \xmm. Left panel shows the first double burst observed by \nustar only. In the right hand panel we show the \xmm and \nustar data of the second occurrence of a double burst. In the second case the first burst is only visible in the unfiltered \nustar data. Note that since no barycentric correction could be applied to unfiltered \nustar data we shifted the times for clarity.}
    \label{fig:doubleburst}
\end{centering}
\end{figure*}

\subsubsection{Time Resolved Spectroscopy with \xmm and \nustar}
\label{sec:spectroscpy}

The effective area of \xmm and \nustar allows us to monitor the X-ray spectral evolution during the bursts. For this purpose, we followed a dynamic time-binning scheme; we extracted X-ray spectra in 2~s intervals, but the integration time was iteratively increased by a factor of $\sqrt{2}$ whenever the net count rate in the 1.0--10~keV band fell below 150 counts, ensuring sufficient statistics for spectral fitting. A 100~s pre-burst spectrum were also extracted to characterize the persistent emission. For each time bin, instrumental response files were generated using \texttt{rmfgen} and \texttt{arfgen}, and spectra were grouped to have a minimum of 25 counts per bin with \texttt{specgroup}.

Time-resolved spectral fitting of \xmm EPIC-pn burst data from \src was performed using the Sherpa \citep{2024ApJS..274...43S} fitting package. The pre-burst (persistent) emission was characterized by fitting the 100~s pre-burst spectrum with an absorbed power-law model. Following the spectral analysis of the persistent emission, we also tried applying the \emph{comptt} model, however because of the low count rate of the source, in most cases the best fit parameters of the model could not be constrained. For the burst spectra the parameters of the pre-burst spectral model was held fixed. Because some of the bursts occur very close to X-ray dips we allowed the Hydrogen column density to vary throughout the observation. Results of the best fit pre-burst spectral parameters and the 1.0--10.0~keV fluxes are given in \autoref{tab:pre_burst_params}. We note that the use of a simple power-law model instead of the \texttt{comptt} model introduces an artificial increase in the inferred Hydrogen column density values. 

We fitted the burst spectra by adding a blackbody component (\emph{bbodyrad}) to the persistent emission model. The resulting spectral evolution for each burst is shown in \autoref{fig:ts_res} and \autoref{fig:ts_res2}. These figures show the spectral evolution of the best-fit parameters, their 1$\sigma$ uncertainties, and the bolometric fluxes with Monte Carlo errors, together with the resulting  $\chi^2$ values. We present the peak fluxes, and the blackbody parameters at the peak flux moment in \autoref{tab:burst_peaks}.

Note that \xmm bursts \#2, \#4, \#6, and \#7, were also observed with \nustar simultaneously, allowing us to also extract \nustar FPMA/FPMB (3--20~keV) spectra and perform a simultaneous fit. During these fits we added the same constant offset value we found during the persistent state analysis.  

\begin{table*}
\centering
\caption{Best fit pre-burst spectral parameters as obtained from \xmm. For bursts \#5, \#7, \#9, and \#10 we provide the results from the simultaneous fits with \nustar.}
\label{tab:pre_burst_params}
\begin{tabular}{ccccc}
\hline
Burst ID & $N_H$ ($10^{22}$ cm$^{-2}$) & $\Gamma$ & Flux$^*$ & $\chi^2$/dof \\
\hline
4 & 5.78$\pm$0.76& 1.44$\pm$0.16& 0.12$\pm$0.04& 18.6/26 \\
5 & 7.20$\pm$0.79 &1.60$\pm$0.11&0.17$\pm$0.04& 51.3/68 \\
6 & 5.49$\pm$0.80 & 1.47$\pm$0.19 & 0.11$\pm$0.05 & 28.8/25 \\
7 & 6.09$\pm$0.68 &1.58$\pm$0.11 & 0.15$\pm$0.04 & 68.2/67 \\
8 & 4.76$\pm$0.70 & 1.34$\pm$0.20 & 0.09$\pm$0.04 & 19.9/22 \\
9 & 5.46$\pm$0.59 & 1.58$\pm$0.10 & 0.16$\pm$0.04 & 92.1/73 \\
10 & 6.24$\pm$0.62 & 1.56$\pm$0.10 & 0.16$\pm$0.04 & 53.1/68 \\
11 & 5.23$\pm$0.81&1.59$\pm$0.18 & 0.11$\pm$0.04 & 17.0/24 \\
12 & 5.10$\pm$0.90&1.31$\pm$0.20 & 0.11$\pm$0.05 & 17.1/24 \\
\hline
\end{tabular}
\footnotesize{\\ $^*$ Pre-burst fluxes are calculated for the 1--10~keV energy range for \xmm only bursts. For bursts \#5, \#7, \#9, and \#10 the fluxes are given for 1--20~keV energy range. All the flux values are in units of $10^{-9}$ erg/s/cm$^2$.}
\end{table*}

\begin{table}
\centering
\caption{Peak fluxes of the bursts detected with \xmm and the spectral parameters at the peak flux moment. For \xmm bursts \#5, \#7, \#9, and \#10 we provide the results from the simultaneous fits with \nustar.}
\label{tab:burst_peaks}
\begin{tabular}{cccc}
\hline
Burst ID & Flux$^{*}$ & $kT$ & Normalization \\
&   & (keV) & R$^2/D_{10 \rm{kpc}}^2$ \\
\hline
4 & 4.01$\pm$0.59&2.74$\pm$0.36& 8.2$\pm$2.6\\
5 & 3.92$\pm$0.23&2.42$\pm$0.12&11.9$\pm$1.9\\
6 & 4.17$\pm$0.43&2.52$\pm$0.22&11.2$\pm$2.6 \\
7 & 2.15$\pm$0.15&1.83$\pm$0.10&18.6$\pm$3.4\\
8 & 3.81$\pm$0.36&2.34$\pm$0.18&13.3$\pm$2.8\\
9 & 3.31$\pm$0.22&2.69$\pm$0.14&6.9$\pm$1.1 \\
10 & 3.81$\pm$0.23&2.85$\pm$0.15&6.6$\pm$1.0\\
11 & 4.56$\pm$0.54&2.97$\pm$0.31&7.1$\pm$1.8\\
12 & 4.33$\pm$0.47&2.65$\pm$0.25&9.7$\pm$2.3\\
\hline
\end{tabular}
\footnotesize{\\ $^*$ Unabsorbed bolometric fluxes in units of $10^{-9}$ erg/s/cm$^2$.}
\end{table}

\subsubsection{Search for evidence of coronal cooling}
\label{sec:coronalcooling}

To search for any evidence of cooling of corona in \src, we analysed the 
$40$--$79$~keV lightcurve from the \nustar observation around each detected 
X-ray burst. Given the low count rates at higher energies, we stacked bursts 
with similar profiles. To select the appropriate bursts for stacking, we computed normalised burst profiles by subtracting the pre-burst count rate (measured in a 150~s window prior to onset) and dividing by the peak net rate, with onset aligned to the first bin where the normalised $3$--$79$~keV rate exceeded $15\%$ of peak for at least two consecutive 1~s bins. Inspection of the aligned profiles expectedly revealed that Bursts \#2 and \#7 -- the secondary bursts noted above -- have significantly 
shorter $e$-folding times ($\tau\approx10$~s; see also 
\autoref{tab:nustar_burst_values}), and were excluded. The remaining five bursts 
(B1, B3, B5, B8, B10) form a morphologically homogeneous group with 
$\tau\approx20$~s and were used for the Compton cooling search.

We performed a Bayesian model comparison between a constant hard X-ray rate 
($\mathcal{M}_0$) and a model including a rectangular flux dip of duration 
$\Delta t_{\rm dip}$ beginning at burst onset ($\mathcal{M}_1$), where the 
rate drops from $C_{\rm pre}$ to $C_{\rm dip}\leq C_{\rm pre}$:

\begin{equation}
  \lambda(t) =
  \begin{cases}
    C_{\rm pre} & t < 0\ \text{or}\ t > \Delta t_{\rm dip}, \\
    C_{\rm dip} & 0 \leq t \leq \Delta t_{\rm dip}.
  \end{cases}
\end{equation}

Both rates were assigned uniform priors on $[0,\,C_{\rm max}]$ with 
$C_{\rm max}=10.92$~cts~s$^{-1}$. Marginal likelihoods were computed using 
\textsc{dynesty} \citep[v3.0;][]{speagle2020,koposov2022} with $n_{\rm live}=400$ 
live points. We tested dip durations of $\Delta t_{\rm dip}=5$, $10$, and $20$~s; 
the $5$~s case is physically most relevant as it targets the burst rise, when the 
initial surge of seed photons is expected to quench the corona most rapidly. 
For this case:

\begin{equation}
  \ln\mathcal{Z}_0 = -1996.24\pm0.13, \qquad \ln\mathcal{Z}_1 = -1999.39\pm0.16,
\end{equation}

\noindent giving $\ln B_{10} = -3.15\pm0.20$ ($B_{10}=0.043$; $1\sigma$ interval 
$[0.035,\,0.052]$), constituting strong evidence against a cooling dip on the 
Jeffreys scale \citep{jeffreys1961,kass1995}. Results for the other tested 
durations and energy ranges (including extending the band to $30$~keV) were 
consistent.

\subsubsection{Upper Limits on Cooling Depth}
\label{sec:cooling_upper_limits}

\begin{figure}
  \centering
  \includegraphics[width=\columnwidth]{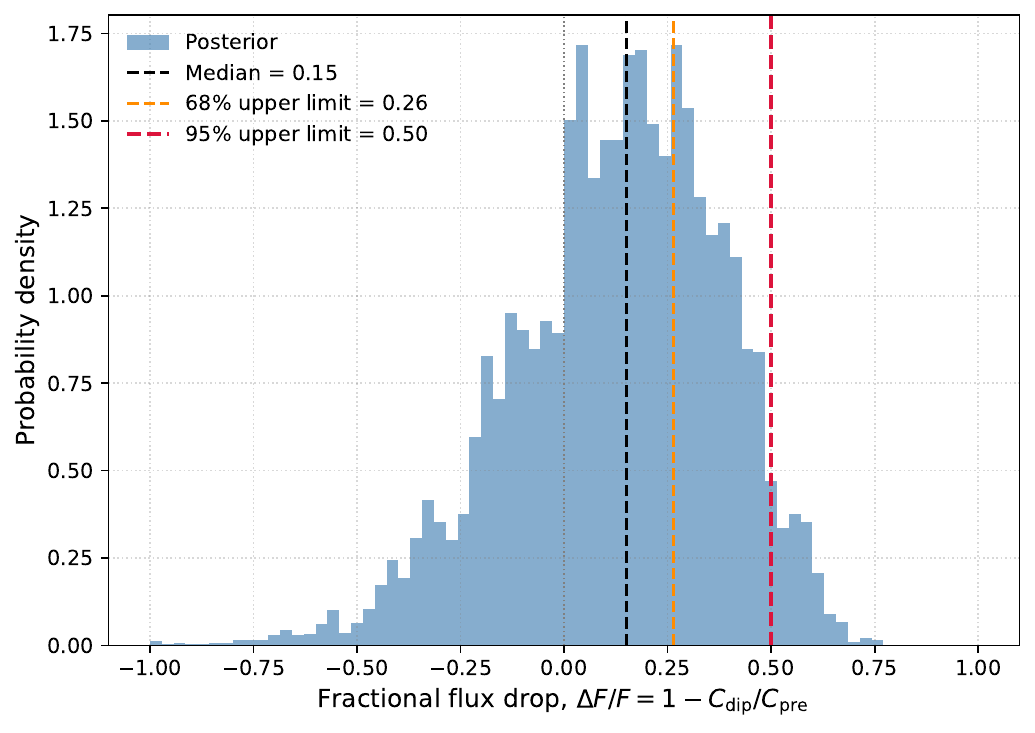}
  \caption{Marginalised posterior of the fractional hard-band flux drop 
    $f_{\rm drop}=1-C_{\rm dip}/C_{\rm pre}$ under $\mathcal{M}_1$ 
    ($\Delta t_{\rm dip}=5$~s). Vertical lines mark the $68\%$ and $95\%$ 
    credible upper limits ($f_{\rm drop}<0.26$ and $<0.50$, respectively). 
    The broad, flat posterior indicates no evidence for flux suppression.}
  \label{fig:cooling_posterior}
\end{figure}

Although no cooling signal is detected, the $\mathcal{M}_1$ posterior constrains 
the depth of any undetected dip via the fractional flux drop 
$f_{\rm drop}\equiv1-C_{\rm dip}/C_{\rm pre}$. The posterior~
(\autoref{fig:cooling_posterior}) yields a median of $f_{\rm drop}=0.15$, 
with a $95\%$ credible upper limit of $f_{\rm drop}<0.499$. We therefore conclude 
that the $40$--$79$~keV flux of \src did not decrease by more than $50\%$ at the 
onset of its Type~I bursts, at $95\%$ credibility, for dip durations up to 
$\sim20$~s.

\subsection{X-ray Dips} 
\label{sec:dips}

\begin{figure*}
    \centering
    \includegraphics[scale=0.7]{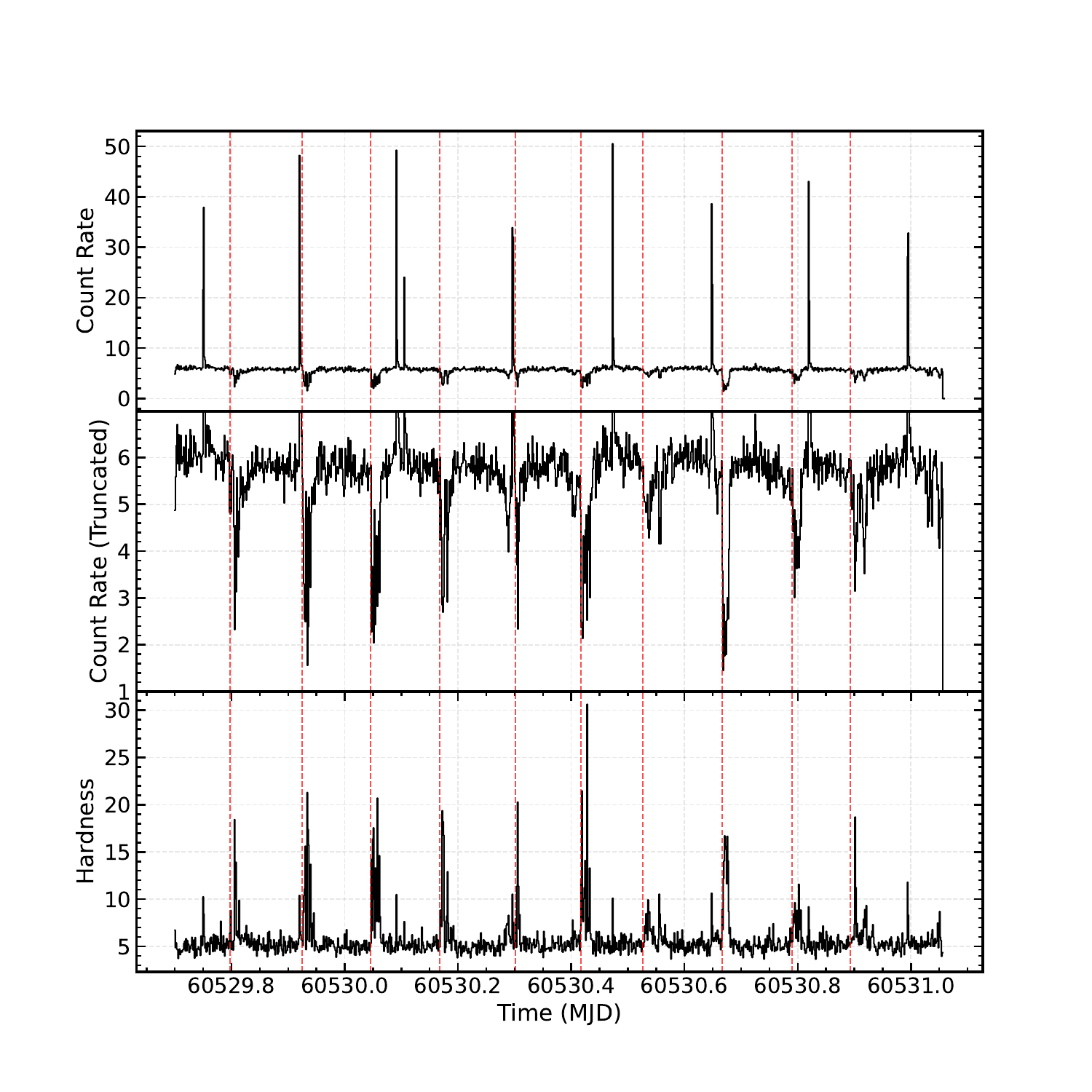}
    \caption{0.6--10.0~keV lightcurve of \src obtained with \xmm with a binning of 100~s. Upper panel shows the full lightcurve, while the middle panel shows a limited range in count rate to emphasize the X-ray dips. The lower panel shows the hardness ratio defined as the ratio of counts detected in the 2.5--10 / 0.6--2.5 keV. During the dips the hardness increases significantly and makes the identification of the dips easier.}
    \label{fig:dips}
\end{figure*}

X-ray dips observed from several LMXBs are thought to be caused by obscuring colder material in the line of sight, for this reason, it is expected that during an X-ray dip the brightness should drop, while the spectrum hardens. Therefore, we searched for dips using 0.6--10.0~keV X-ray light curves and hardness ratios. We defined the hardness ratio as the rate of counts detected in the 2.5--10.0~keV band to the counts detected in the 0.6--2.5~keV band.  We show the \xmm lightcurve and the time evolution of the hardness ratio in \autoref{fig:dips}, we also show zoomed in views of each X-ray dip in \autoref{fig:dipszoom}. In both figures the time resolution is 100~s. In total we detect 10 X-ray dips. Start times, end times, durations, the minimum count rates and the maximum hardness ratio observed are given in \autoref{tab:dip_times}.  In \autoref{tab:dip_times} we also provide the time difference between the maximum hardness ratio observed during a dip and the time minimum count rate is seen. While typically the times match, within the time resolution of our lightcurves, in three cases there is significant delay in hardness by as much as 10 minutes. This observed delay between the minimum count rate and the maximum hardness ratio may suggest that the absorbing material is not uniformly distributed, with denser clumps trailing the leading edge of the bulge structure \citep[see, e.g.,][]{2004MNRAS.348..955C,2006A&A...445..179D}.

In terms of dip durations, typical value we infer is half an hour, in Dip \#5 we see that the duration decreases to about 10 minutes but this dip happens just after a burst and before the burst a slight decrease in the count rate is evident hinting that the dips started just before the burst. Also, the Dips \#8 and \#9 only last about 20 minutes, but again in both cases we see that a thermonuclear burst happens just before or just after the X-ray dip.

The time in between the dips is also very regular. A linear fit to the time intervals gives 2.942 $\pm$ 0.011 hours or 176.54 $\pm$ 0.65 min for the periodicity. This periodicity is in very much agreement with earlier orbital periods found for \source \citep{1995AJ....110.1292S,Parmar_7600607,2005MNRAS.359.1336C,Gambino_7600607,2009A&A...500..873B}.
\begin{table*}
\centering
\caption{Properties of all the X-ray dips detected in the \xmm data.}
\begin{tabular}{ccccccccc}
\hline 
Dip & Start Time & End Time & Duration & Time Between & Min. Count Rate & Max. Hardness Ratio & Hardness Delay & \\
ID & MJD &MJD & (hours) & (hours) &(c~s$^{-1}$) & & (minutes) \\
\hline
1 &  60529.797 & 60529.820& 0.5278& -- &2.33 $\pm$ 1.53 & 18.4 & -- \\
2 & 60529.925 & 60529.950& 0.5556& 3.10&1.56 $\pm$ 1.25 & 21.3 & -- \\
3 & 60530.046 & 60530.065& 0.444& 2.83 &2.04 $\pm$ 1.43 & 20.7 & 10.0\\
4 & 60530.168 & 60530.190& 0.500& 2.97&2.69 $\pm$ 1.69 & 19.4& 1.6\\
5 & 60530.301 & 60530.310& 0.1667&3.03 &2.34 $\pm$ 1.53 & 20.2& --\\
6 & 60530.417 & 60530.440& 0.5278& 2.96&2.14 $\pm$ 1.46 & 30.6 & 11.66\\
7 & 60530.527 & 60530.560& 0.7500& 2.75&4.15 $\pm$ 2.04 & 10.5 & -- \\
8 & 60530.666 & 60530.680& 0.3056& 3.11&1.45 $\pm$ 1.20 & 16.7 & 3.33\\
9 & 60530.790 & 60530.805& 0.3333& 2.99&3.00 $\pm$ 1.73 & 11.6 & 11.66\\
10 & 60530.893 &  60530.925& 0.7500& 2.68&3.15 $\pm$ 1.77 & 18.7 & -- \\ 
\hline
\end{tabular}
\label{tab:dip_times}\\
\end{table*}

In order to quantify the spectral variation during the dips we extracted X-ray spectra of the data within the time intervals given in \autoref{tab:dip_times}. Similar to the spectral analysis of the persistent times, we fit the X-ray spectra with an absorbed \texttt{comptt} model assuming a constant fixed offset in between the EPIC pn and \nustar FPMA and FPMB. The offset value is fixed to  the value inferred from the spectral analysis of the persistent phases. We also fixed the plasma temperature of the \texttt{comptt} model in our fits.

Since X-ray dips in these systems are thought to be due to colder material absorbing the X-ray emission from the central region of the accretion disk, in our fits we allowed the hyrdogen column density to vary, to see if it has any correlation with the dip properties. We give our best fit results in \autoref{tab:dip_fits_comptt}. 

\begin{table*}
\centering
\caption{Best-fit spectral parameters for the 4U\,1323$-$62 dip spectra. }
\label{tab:dip_fits_comptt}
\setlength{\tabcolsep}{4pt}
\begin{tabular}{lcccccc}
\hline\hline
Dip &
$N_{\rm H}$ &
$T_0$ &
$\tau_{\rm p}$ &
$F_{0.5\text{--}10}$$^{*}$ &
$F_{3\text{--}30}$$^{*}$ &
$\chi^{2}/\nu$ \\
 &
($10^{22}$\,cm$^{-2}$) &
(keV) &
 &
 &
 &
 \\
\hline
Dip\,1 & $3.01^{+0.21}_{-0.20}$ & $1.05^{+0.03}_{-0.03}$ & $2.75^{+0.10}_{-0.10}$ & $8.3^{+0.1}_{-0.1}$ & $16.9^{+0.5}_{-0.5}$ & 195.3/198 \\[2pt]
Dip\,2 & $3.33^{+0.24}_{-0.22}$ & $1.03^{+0.04}_{-0.03}$ & $3.01^{+0.09}_{-0.09}$ & $7.6^{+0.1}_{-0.1}$ & $16.7^{+0.4}_{-0.4}$ & 219.4/229 \\[2pt]
Dip\,3 & $3.86^{+0.32}_{-0.29}$ & $1.11^{+0.04}_{-0.04}$ & $2.85^{+0.10}_{-0.10}$ & $7.1^{+0.1}_{-0.1}$ & $15.9^{+0.4}_{-0.4}$ & 227.1/192 \\[2pt]
Dip\,4 & $3.65^{+0.25}_{-0.23}$ & $0.97^{+0.03}_{-0.03}$ & $2.93^{+0.07}_{-0.07}$ & $8.1^{+0.1}_{-0.1}$ & $17.8^{+0.4}_{-0.4}$ & 407.9/370 \\[2pt]
Dip\,5 & $3.45^{+0.41}_{-0.36}$ & $1.12^{+0.05}_{-0.05}$ & $2.58^{+0.11}_{-0.11}$ & $7.7^{+0.2}_{-0.2}$ & $16.7^{+0.5}_{-0.5}$ & 233.8/195 \\[2pt]
Dip\,6 & $3.39^{+0.25}_{-0.24}$ & $1.11^{+0.04}_{-0.04}$ & $2.62^{+0.13}_{-0.14}$ & $7.3^{+0.1}_{-0.1}$ & $18.8^{+0.7}_{-0.7}$ & 212.8/204 \\[2pt]
Dip\,7 & $3.11^{+0.17}_{-0.16}$ & $0.99^{+0.03}_{-0.03}$ & $2.67^{+0.08}_{-0.08}$ & $8.9^{+0.1}_{-0.1}$ & $17.9^{+0.4}_{-0.4}$ & 354.1/337 \\[2pt]
Dip\,8 & $5.58^{+0.71}_{-0.63}$ & $1.17^{+0.06}_{-0.06}$ & $3.15^{+0.14}_{-0.14}$ & $5.7^{+0.2}_{-0.2}$ & $14.3^{+0.5}_{-0.5}$ & 284.4/192 \\[2pt]
Dip\,9 & $4.45^{+0.44}_{-0.39}$ & $0.98^{+0.05}_{-0.05}$ & $2.86^{+0.09}_{-0.09}$ & $8.0^{+0.2}_{-0.2}$ & $18.2^{+0.4}_{-0.4}$ & 293.0/286 \\[2pt]
Dip\,10 & $3.17^{+0.18}_{-0.17}$ & $1.00^{+0.03}_{-0.03}$ & $2.88^{+0.07}_{-0.07}$ & $8.5^{+0.1}_{-0.1}$ & $17.3^{+0.3}_{-0.3}$ & 380.4/376 \\
\hline
\end{tabular}
\footnotesize{\\ $^*$ Fluxes are unabsorbed and given in units of $10^{-11}$\,erg\,cm$^{-2}$\,s$^{-1}$. The 0.5--10\,keV flux is from \xmm EPIC-pn; the 3--30\,keV flux is from \nustar~FPMA.}
\end{table*}

\section{Discussion}
\label{sec:conclusion}

In this study we report the latest X-ray observations of the dipping clocked X-ray burster \src. Within the non-bursting, non-dipping time intervals we detected a QPO at $\sim$0.898~Hz, which is consistent but on the lower end of the previous detections at around 0.97 -- 1.4 Hz reported from this source  \citep{1999ApJ...511L..41J,2000ApJ...542L.111T,2010MNRAS.407.1895B,Bhulla_2020}. Similar low-frequency QPOs have been observed in other high-inclination dipping LMXBs, such as EXO 0748-676 \citep{1999ApJ...516L..91H}, suggesting a common geometric origin linked to the high inclination of these systems.

Spectral analysis of the non-bursting, non-dipping times in the 2.0 -- 50.0~keV range shows that the spectra can be fit with an absorbed \texttt{comptt} model. Given the dipper nature of the source we also attempted to fit the data with a relativistic reflection model using the \texttt{relxillCp}. However, due to the low count rate of the source, the model could not yield adequately constrained parameters. These reflection models also did not resolve the residuals at lower energies as well. 

\begin{figure}
\centering
\includegraphics[width=1.0\linewidth]{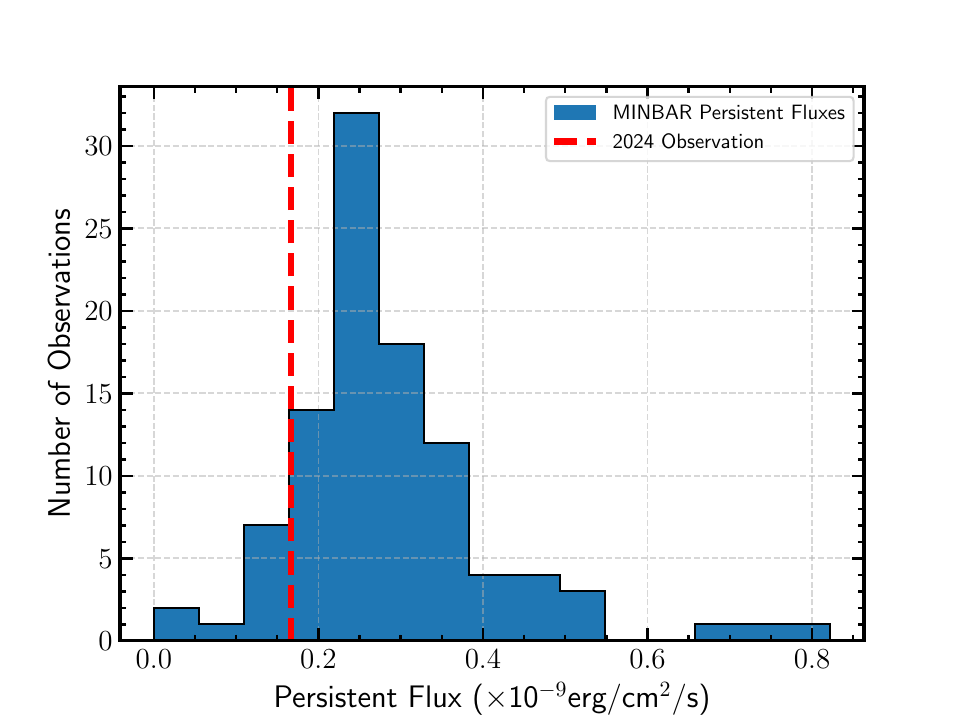}
\caption{Comparison of the archival persistent flux values reported from the MINBAR catalog \citep{2020ApJS..249...32G} using absorbed \texttt{comptt} model in the 3--25~keV range with the flux we measure during our 2024 observation (red dashed line).}
\label{fig:minbar_pers_hist}
\end{figure}

The source flux we infer from our spectral fits reveal that the source was in a significantly low state compared to earlier observations of the source \cite[see e.g.,][]{Boirin_7600607, Bhulla_2020}. MINBAR catalog \citep{2020ApJS..249...32G} reports the pre-burst persistent flux of the source for 99 different bursts in between 1997 -- 2012. For comparison with previous persistent fluxes reported for \source, we also calculated 3.0 -- 25.0~keV flux as (1.66$\pm0.01)\times10^{-10}$~\fluxcgs. In \autoref{fig:minbar_pers_hist} we show a comparison of the flux we measure using \nustar with the archival RXTE/PCA fluxes reported in the catalog. 

\cite{2005MNRAS.359.1336C} compared the source luminosity and burst rate among many observations. We also compiled a list of some of the historic observations of \source with mean burst rates (see \autoref{tab:obs_log}). We note that in \autoref{tab:obs_log} we provide the net effective exposure times to calculate propoer burst rates. \cite{2005MNRAS.359.1336C} noted that as the source luminosity increases the burst rate also increase. As mentioned above during our observation the source flux is at one of the lowest levels and hence the burst rate is also at one of the lowest. Our spectral analysis revealed that the source luminosity in 2024 is $\approx 1.2 \times 10^{36} \ \text{erg/s}$ in the 1--10~keV range assuming a distance of 10~kpc, which is almost a factor of 5 lower than the 2003 observation with \xmm. We note that the distance to the source was later estimated to be 4.2~kpc \citep{Gambino_7600607}; however, we adopted 10~kpc for easier comparison with previous studies. 

\begin{table*}
\centering
\caption{The observation summary of \source data by different instruments that detect the orbit period. Burst rates are recomputed throughout using the true, good-time-interval-filtered \emph{effective} exposure rather than the elapsed span of the observing campaign (see \S\ref{sec:conclusion} and response to referee); rows marked $^{\dagger}$ were revised from the values reported in the original submission.}
\label{tab:obs_log}
\begin{tabular}{lccccccc}
\hline
Instrument & Observation Time & $N_{\rm burst}$ & Burst Rate & Dips & Orbital Period & Eff. Exposure & Ref. \\
 & & & h$^{-1}$ & & hr & ks & \\
\hline
\textit{EXOSAT}/ME     & 11--13 February 1985 & 6  & 0.200 & 10 & 2.932        & 108.0 & a \\
\textit{RXTE}/PCA      & 25--28 April 1997    & 7  & 0.600 & 7  & 2.45--2.59   & 42.0  & b,g$^{\dagger}$ \\
\textit{BeppoSAX}/MECS & 22--24 August 1997   & 10 & 0.514 & 12 & 2.40--2.57   & 70.0  & h$^{\dagger}$ \\
\textit{XMM-Newton}    & 29 January 2003      & 7  & 0.504 & 5  & 2.97         & 50.0  & d \\
\textit{RXTE}/PCA      & 25 September 2003    & 2  & 0.270 & 4  & 2.45--2.59   & 26.6  & e,g$^{\dagger}$ \\
\textit{SUZAKU}        & 9--10 January 2007   & 5  & 0.147 & 11 & 2.926        & 122.5 & c \\
\textit{CHANDRA}       & 19 December 2011     & 9  & 0.540 & 6  & 2.94         & 60.0  & e \\
\textit{CHANDRA}       & 23 December 2011     & 13 & 0.585 & 8  & 2.94         & 80.0  & e$^{\dagger}$ \\
\textit{LAXPC}         & 16--17 February 2017 & 6  & 0.436 & 2  & $\sim$2.66   & 49.5  & f$^{\dagger}$ \\
\textit{\xmm}          & 7 August 2024        & 9  & 0.239 & 10 & $\sim$2.942  & 135.5 & present \\
\hline
\end{tabular} \\
{\footnotesize
Ref.: $a:$ \cite{Parmar_7600607}; $b:$ \cite{2001A&A...380..494B}; $c:$ \cite{2009A&A...500..873B}; $d:$ \cite{2005MNRAS.359.1336C}; $e:$ \cite{Gambino_7600607}; $f:$ \cite{Bhulla_2020}; $g:$ \cite{2020ApJS..249...32G}; $h:$ \cite{1999A&A...349..495B}.\\}
\end{table*}

Unlike the absorption line features observed earlier by \cite{Boirin_7600607} we detected an asymmetric emission line like a feature at around 6.0~keV. Several studies reported Fe absorption lines around 6.4~keV region \cite[see e.g.,][]{2001A&A...380..494B,Boirin_7600607,2005MNRAS.359.1336C,2006AdSpR..38.2759B}. Typically these features are attributed to highly ionized absorbers, and detected when the source is in a relatively bright state. \cite{2001A&A...380..494B} reported Fe emission line features using RXTE data at around the similar luminosity levels $\sim$3.9$\times$10$^{36}$~erg/s. Unfortunately, the resolution of the RXTE/PCA does not always allow for a search in asymmetry in the line, our \xmm data however reveals a highly asymmetric emission line feature. These observations show us that when the source is in these low burst rate, low luminosity states we detect Fe in emission, while when the source is at higher luminosities with higher burst rates the Fe is observed in absorption.

A total of nine bursts were identified in the \xmm data and nine bursts were detected during the \nustar observation~(see \autoref{fig:all_lc}). Due to an offset in the starting times of the two observations and the inevitable earth occultation of \nustar, only six of the bursts were simultaneously detected with two satellites. Two of these were identified in the \nustar unfiltered data. Within the both data sets we identified two double burst events with separations of 628~s and 1193~s. In relation to the separation times we observe that the peak count rate ratios also decrease by \%62 and \%30, respectively. The detection of multiple such events within a single observation is not unique to these observations. \cite{2001A&A...380..494B, 2009A&A...500..873B, Gambino_7600607, 2008ApJS..179..360G} also report several cases of double bursts observed from this source. In the MINBAR catalog \cite{2020ApJS..249...32G}
there are also examples of double bursts with separation times of several hundred seconds. The fact that the peak intensities reached in the second bursts vary by the time spent in between the two events suggests that  the nuclear fuel was not entirely exhausted during the initial flash, leading to closely spaced secondary ignitions~\citep[see, e.g.,][]{2002ApJ...577..337S,2007A&A...465..559B,2019ApJ...883...61J,2021ApJ...910...37G,2022ApJ...935..154G,2021MNRAS.501..168L,2021ASSL..461..209G}. 

\begin{figure}
    \centering
    \includegraphics[width=1.0\linewidth]{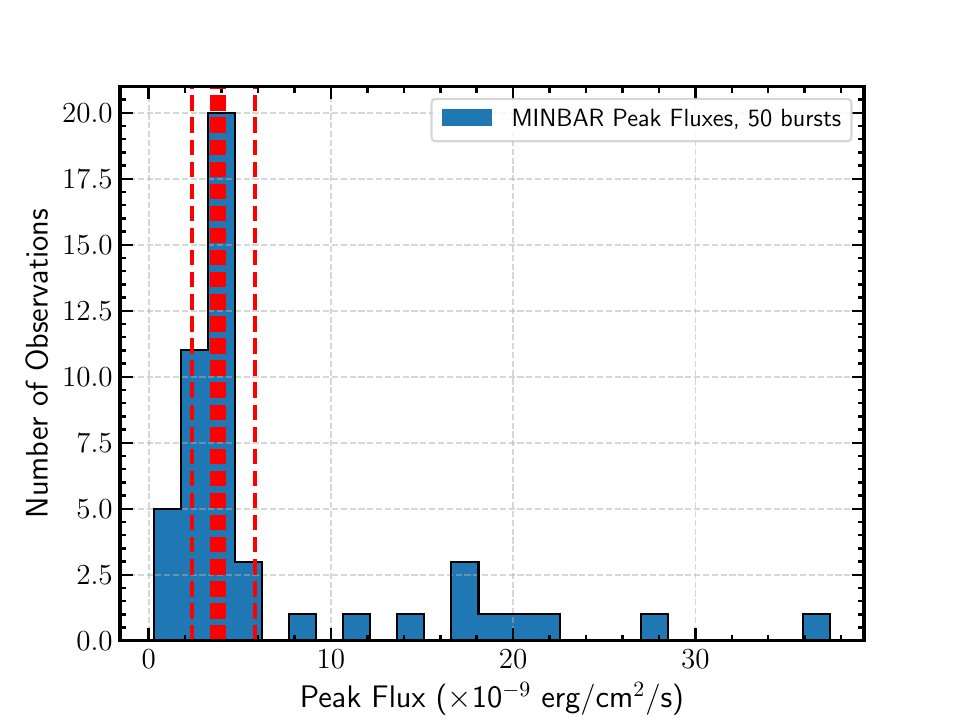}
    \caption{Histogram of all the peak flux measurements reported in the MINBAR archive. Vertical red dashed lines show the peak flux measurements reported in this study.}
    \label{fig:hist_peakflux}
\end{figure}

Time resolved spectroscopy of the bursts allowed us to probe the peak fluxes and the spectral evolution during the bursts. Comparing the peak fluxes with the MINBAR catalog, we found that the peak fluxes of the detected X-ray bursts are in good agreement with the historical peak flux measurements. However, we were not able to detect any of the brightest bursts historically observed.  \autoref{fig:hist_peakflux} show a comparison of the peak fluxes we report.
 
Time resolved X-ray spectroscopy revealed that none of the detected bursts showed any evidence of a photospheric radius expansion. The spectral analysis also did not show a deviation from pure blackbody emission during the bursts. For bursters where the column density in the line of sight is low, soft X-ray data reveal significantly enhanced persistent emission during X-ray bursts, parameterized by the scaling factor $f_a$, \cite[see e.g.,][]{2022MNRAS.510.1577G,2022ApJ...935..154G}. However, it has been observed several times that increase in hydrogen column density damps the soft excess \cite[see e.g.,][]{2021ApJ...910...37G, 2023ApJ...958...55B, 2022ApJ...940...81B}. \source is also affected by the high column density in the line of sight, which is probably the reason for not detecting any soft excess emission.

Despite the low count rate of the source, we searched for any evidence of coronal cooling in the hard X-rays but did not find any evidence for such a correlated variation. The non-detection of Compton cooling can be interpreted within the
\citet{2003A&A...399.1151M} and \citet{keek2016} coronal-evaporation framework.
In that picture, the magnitude of the hard X-ray dip scales approximately as
\begin{equation}
  f_{\rm drop} \approx f_{\rm cor}\,
  \left(1 - e^{-t_{\rm cool}/\tau_{\rm burst}}\right),
\end{equation}
where $f_{\rm cor}$ is the fraction of the persistent $40$--$79$~keV emission that originates in the corona, and $t_{\rm cool} \sim R_{\rm cor}^2/(\kappa_{\rm es} c \ell_{\rm cor})$ is the coronal cooling timescale. The upper limit $f_{\rm drop} < 0.52$ ($95\%$) does not provide a strong constraint on the coronal fraction because at the observed burst fluences and hard-band count rates the photon statistics are limited. Nevertheless, these constraints are consistent with either (i) a corona whose cooling timescale greatly exceeds the burst rise time, (ii) a relatively small coronal fraction of the hard X-ray emission ($f_{\rm cor} \lesssim 0.5$), or (iii) a corona that is already partially quenched at the persistent accretion rate of \src~(an Atoll source accreting near $\dot{m}/\dot{m}_{\rm Edd} \sim 0.01$--$0.05$; \citealt{gierlinski2002}). In addition, it has been shown that the hard X-ray deficit can exhibit an energy dependence attributed to hard X-ray emission from the accretion column \citep{2022ApJ...936L..21C}; however, this is unlikely to apply to  \source given its non-pulsating nature. 
Furthermore, magnetic reconnection in the inner disk region has been proposed as an additional heat source that could partially offset coronal cooling \citep{chen2012}. Finally, the non-detection of coronal cooling in our observations may also suggest that the corona is located further away from the neutron star \citep{2024A&A...685A..71P}, significantly reducing its exposure to the burst emission. In summary, we conclude that the NuSTAR hard X-ray data provide no statistically significant evidence for Compton cooling of the corona during Type~I X-ray bursts in \src.  

\begin{figure}
    \centering
    \includegraphics[width=1.0\linewidth]{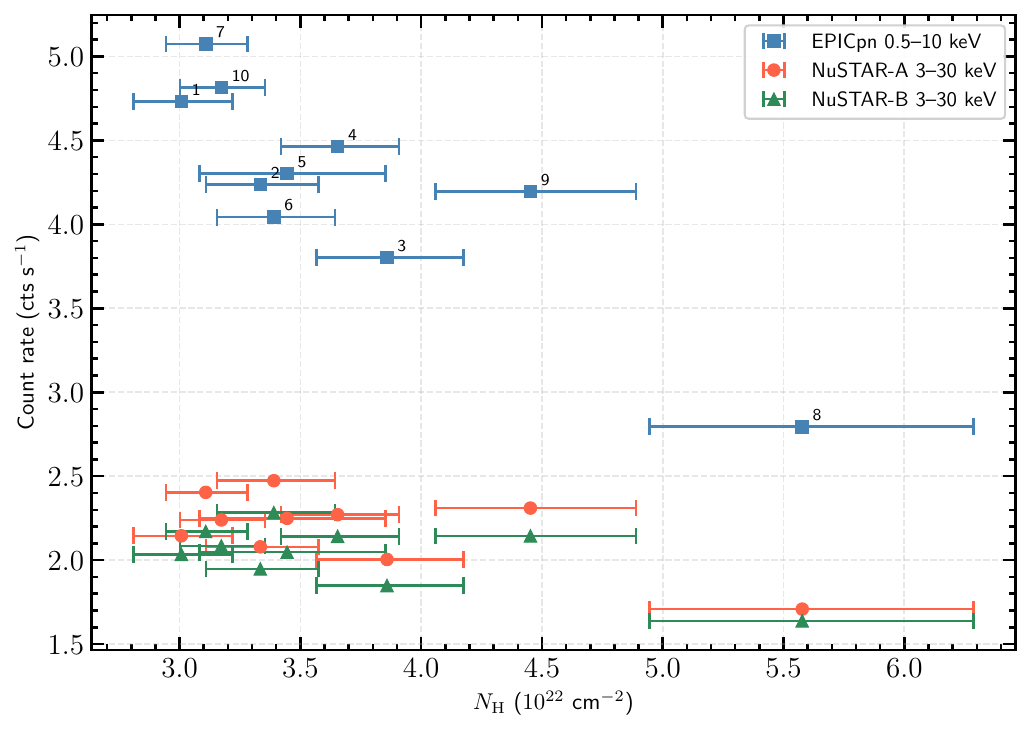}
    \caption{The detected average count rates during the X-ray dips as a function of Hydrogen column density.}
    \label{fig:nh_dip_comp}
\end{figure}

Finally, we detected 10 X-ray dips within our \xmm observation. As can be seen in \autoref{fig:dipszoom} X-ray dips show different structure in each case. While some of the dips seem to be shallow compared to others, in some cases we see the dips to be much cleaner and and deeper. For example Dip \#8 is one of the most significant dips we detected during our observation. We see that the observed average count rate during a dip is correlated with the Hydrogen column density inferred from the X-ray spectra of the dip (see \autoref{fig:nh_dip_comp}) obviously mostly in the soft X-rays. However, in the hard X-ray band, there is no significant correlation with the Hydrogen column density because hard X-rays are not strongly absorbed by the intervening material, the hard X-ray flux remains largely unaffected.

\section*{Acknowledgements}

We thank the anonymous referee whose comments and suggestions improved the paper significantly. T.B., T.G. have been supported by the Scientific Research Projects Coordination Unit of Istanbul University (ADEP Project No: FBA-2026-42278), in part by the Turkish Republic, Presidency of Strategy and Budget project, 2016K121370. T.G. greatly appreciates the hospitality at the School of Physics at the Georgia Institute of Technology, where this work was started. D.R.B. is supported by NASA award 80NSSC24K0212, and NSF grants AST-2307278 \& AST-2407658.
\section*{Data Availability}

The data used in this paper are available in the public archive of HEASARC Web site at \url{https://heasarc.gsfc.nasa.gov/cgi-bin/W3Browse/w3browse.pl}.



\bibliographystyle{mnras}
\bibliography{example} 

\appendix
\section{Lightcurves of Detected X-ray Bursts}

We here show the lightcurves of the detected X-ray bursts with \xmm and \nustar.

\begin{figure*}
\centering
\includegraphics[scale=0.4]{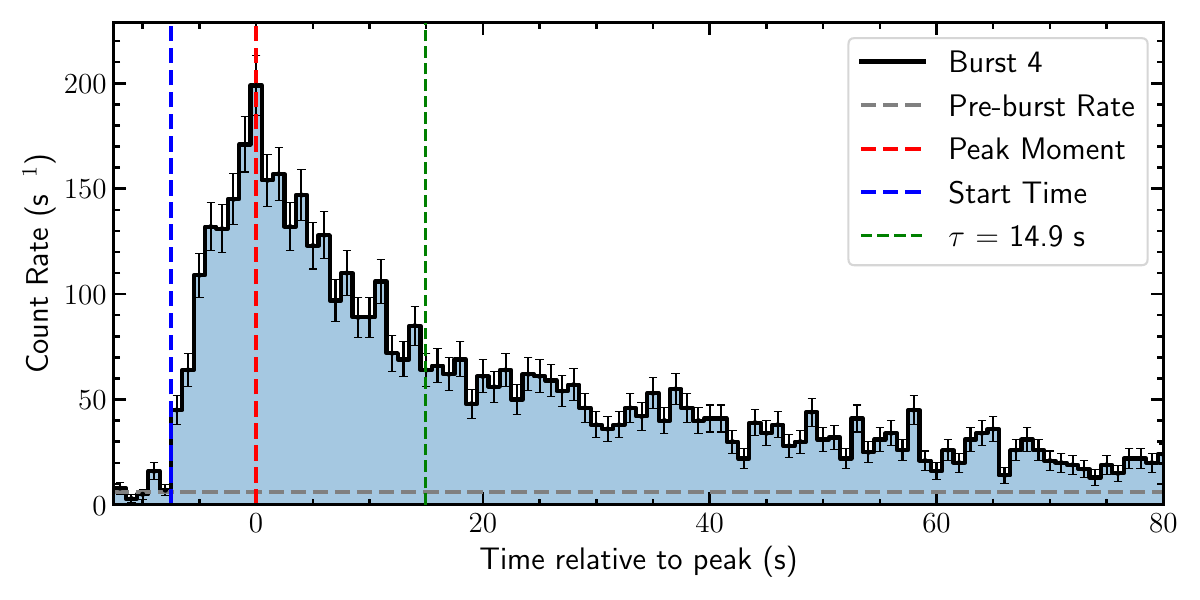} 
\includegraphics[scale=0.4]{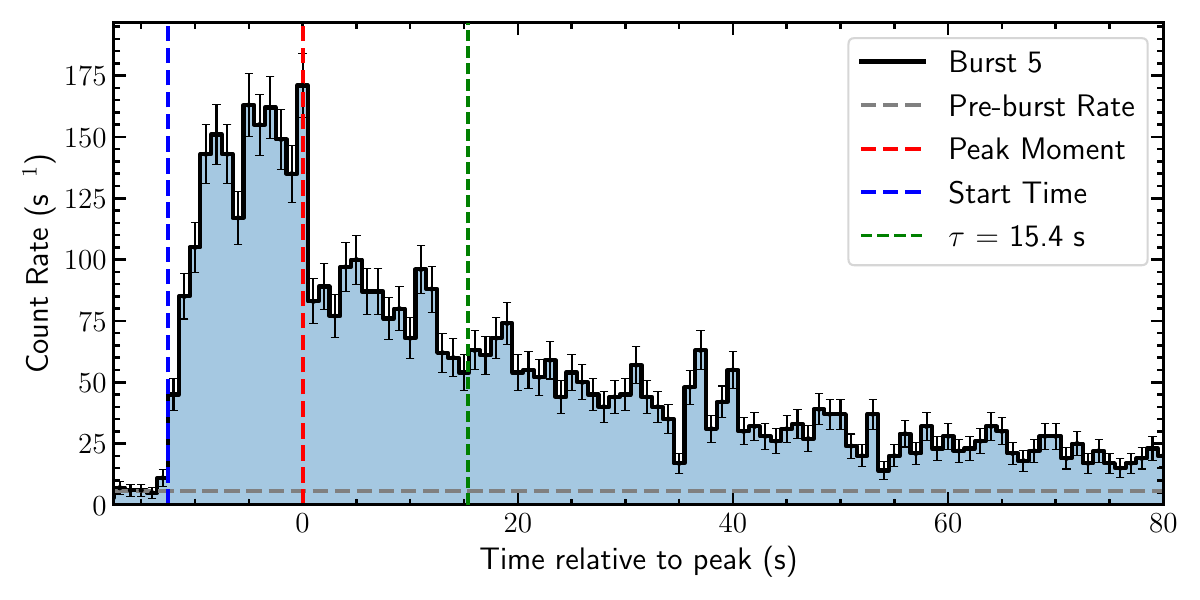}\\ 
\includegraphics[scale=0.4]{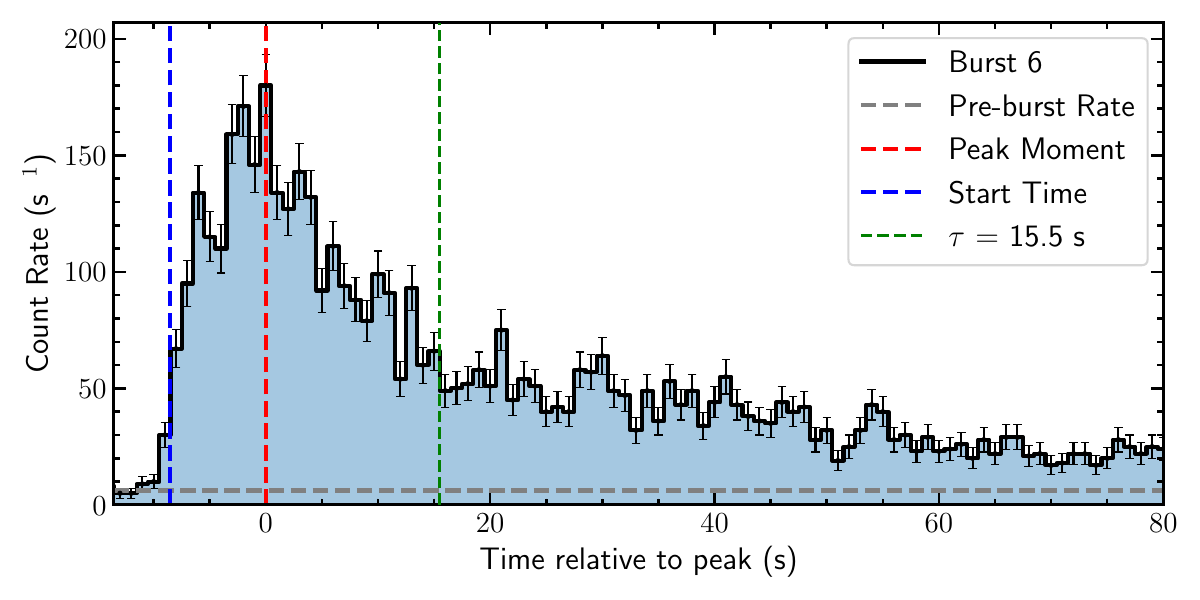} 
\includegraphics[scale=0.4]{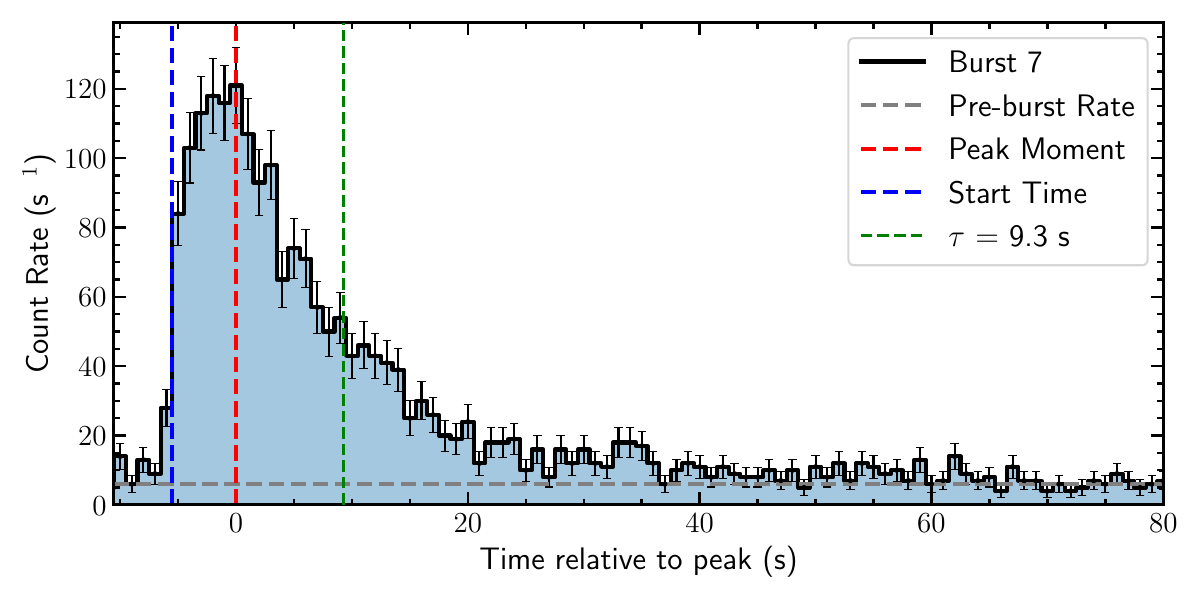}\\ 
\includegraphics[scale=0.4]{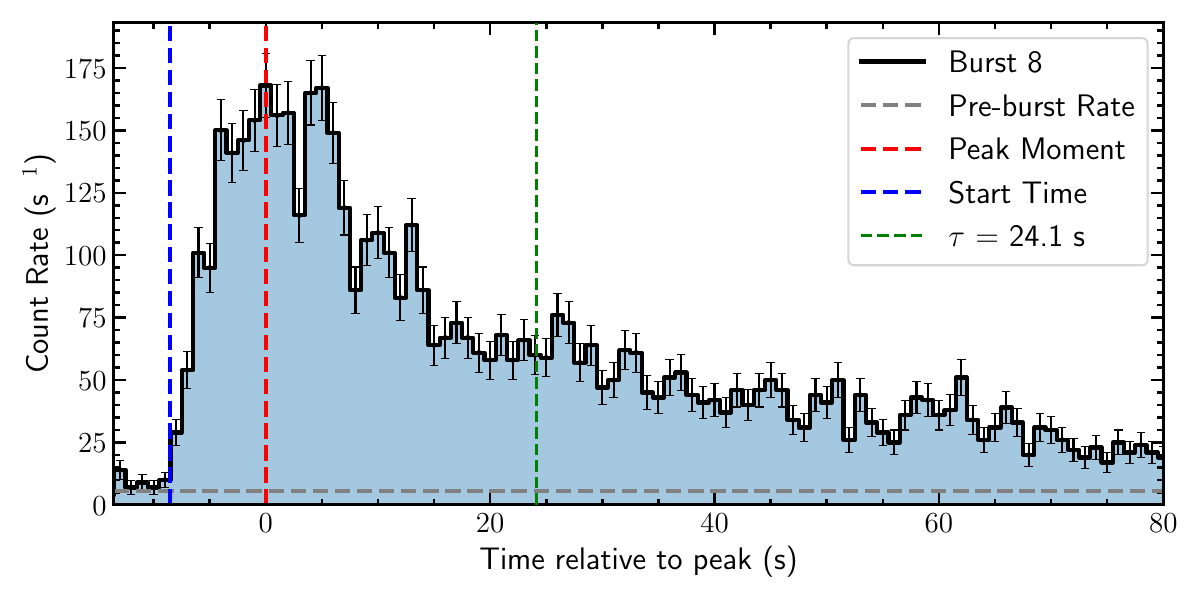} 
\includegraphics[scale=0.4]{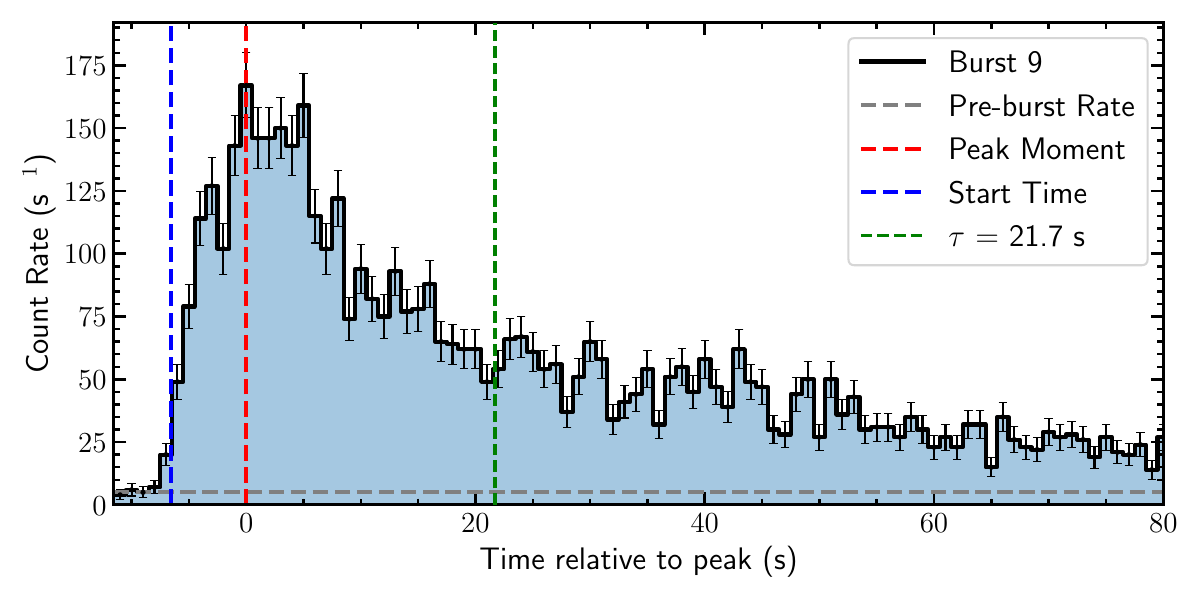}\\ 
\includegraphics[scale=0.4]{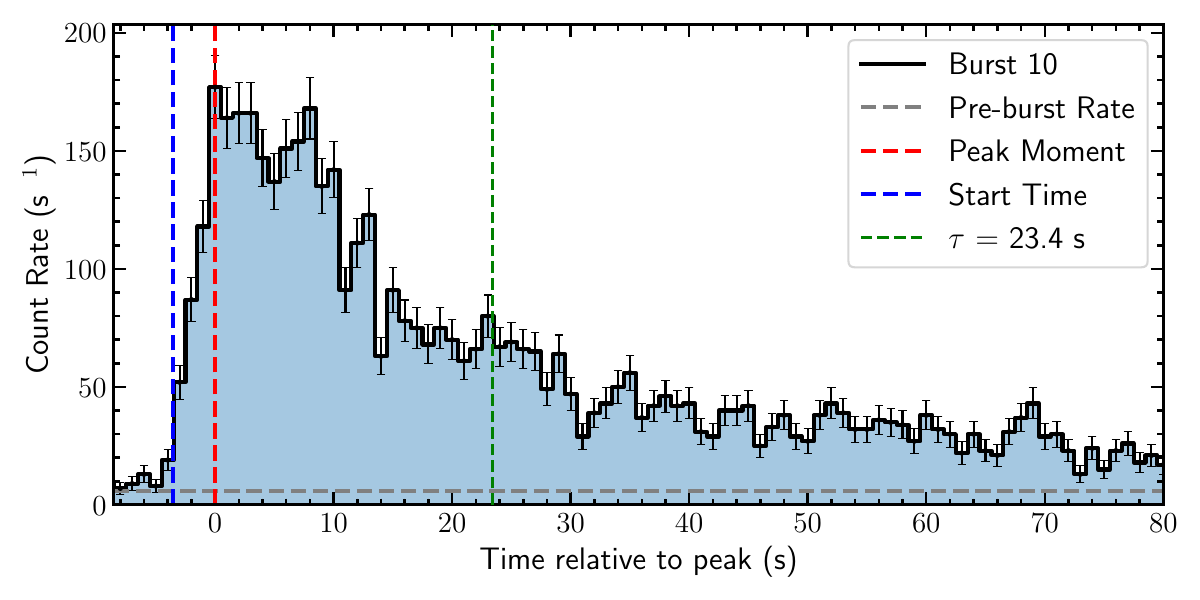}
\includegraphics[scale=0.4]{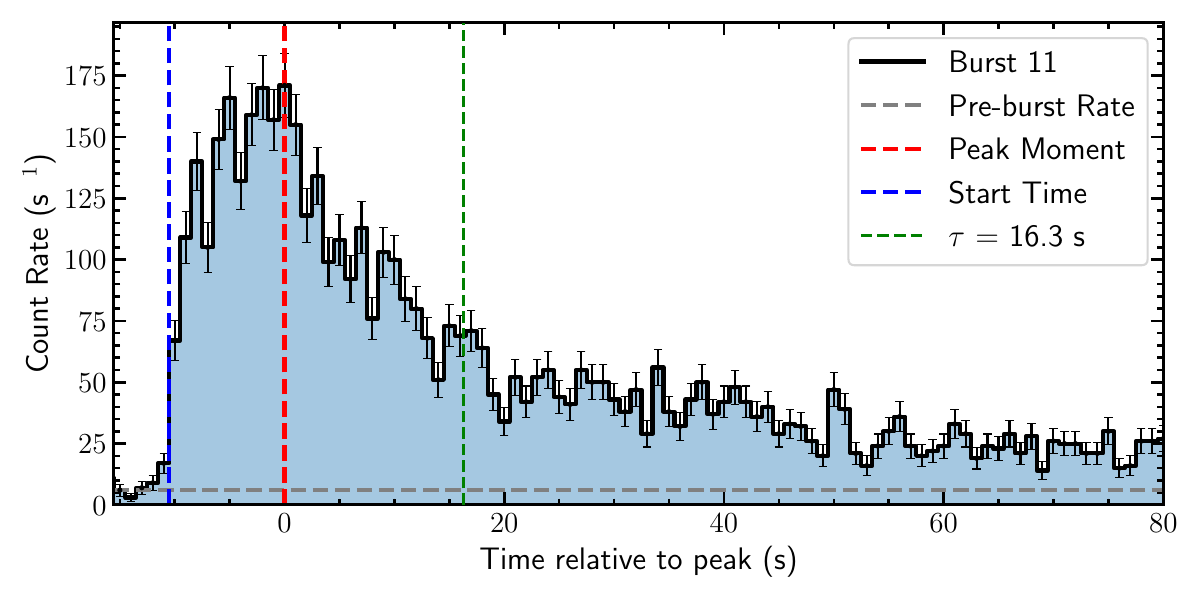}\\ 
\includegraphics[scale=0.4]{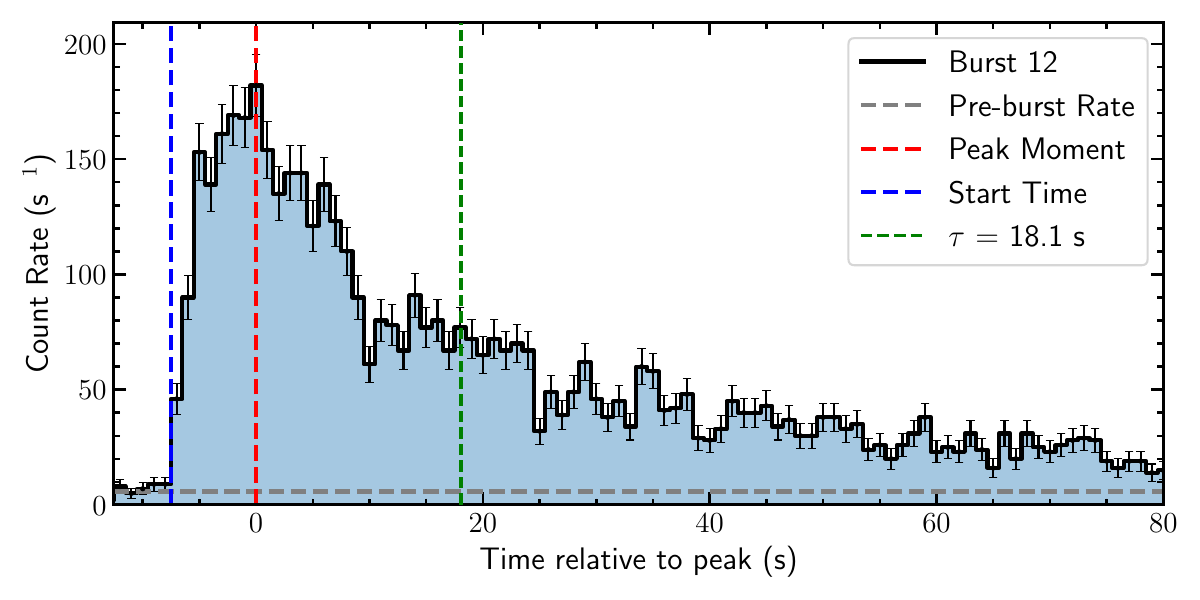}
\caption{0.5--10.0~keV lightcurves of the detected thermonuclear X-ray bursts by \xmm.}
\label{fig:xmmburst_lc}
\end{figure*}

\begin{figure*}
\centering
\includegraphics[scale=0.4]{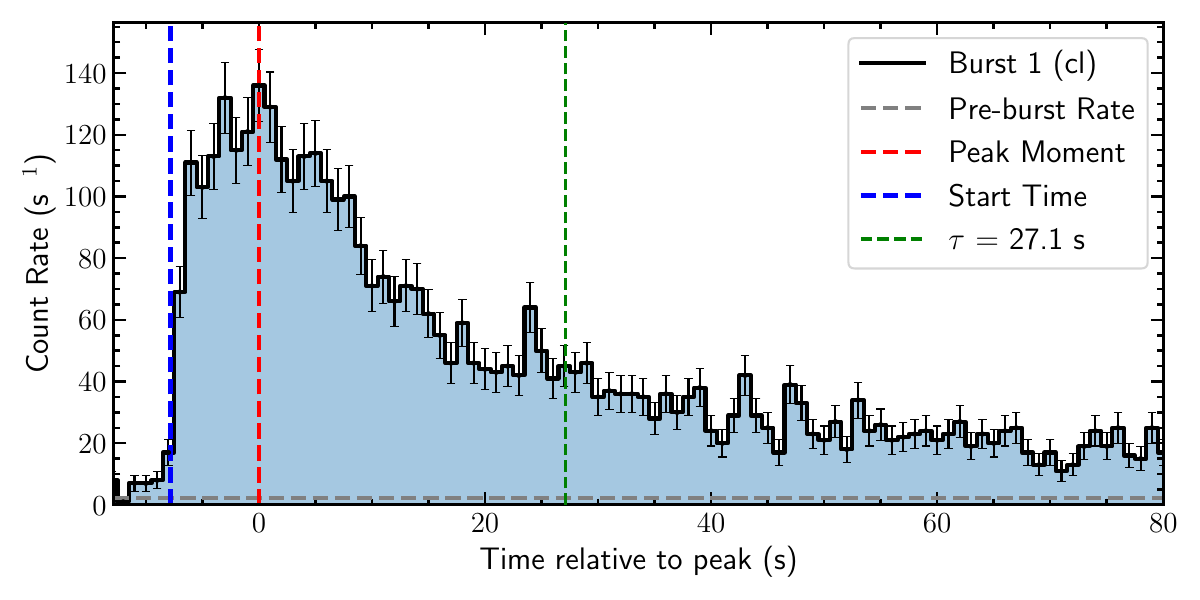}
\includegraphics[scale=0.4]{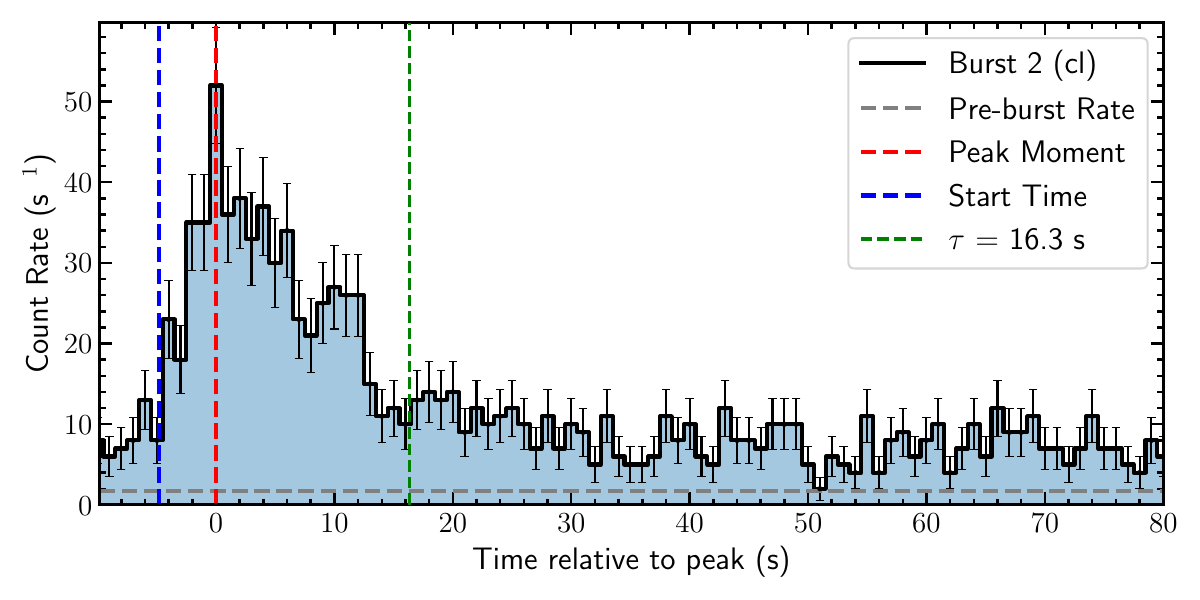} \\
\includegraphics[scale=0.4]{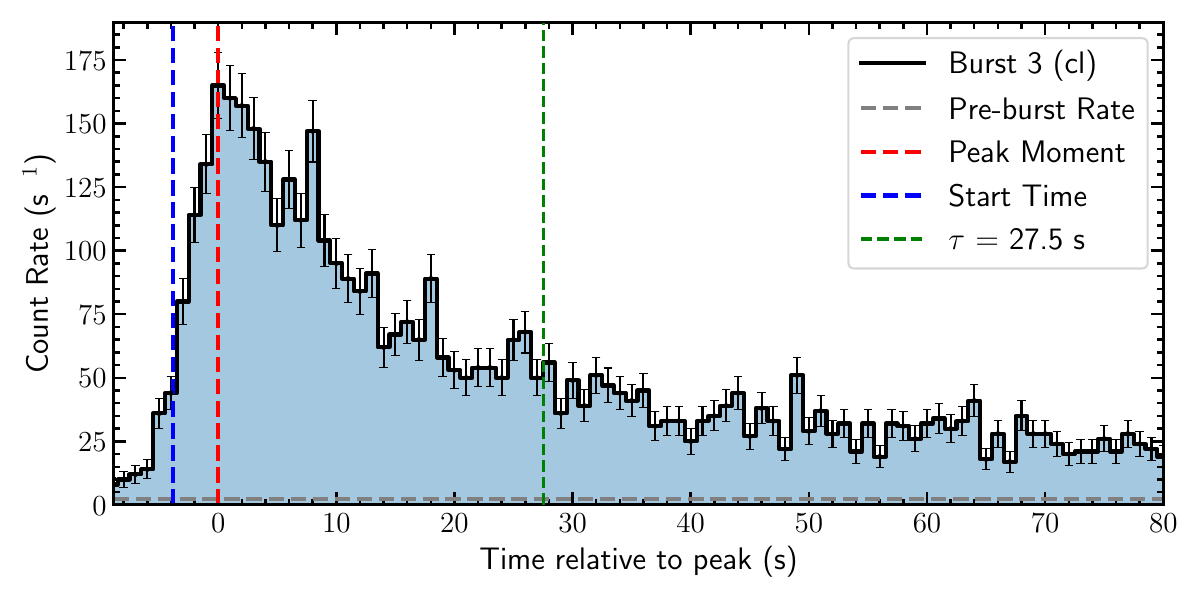} 
\includegraphics[scale=0.4]{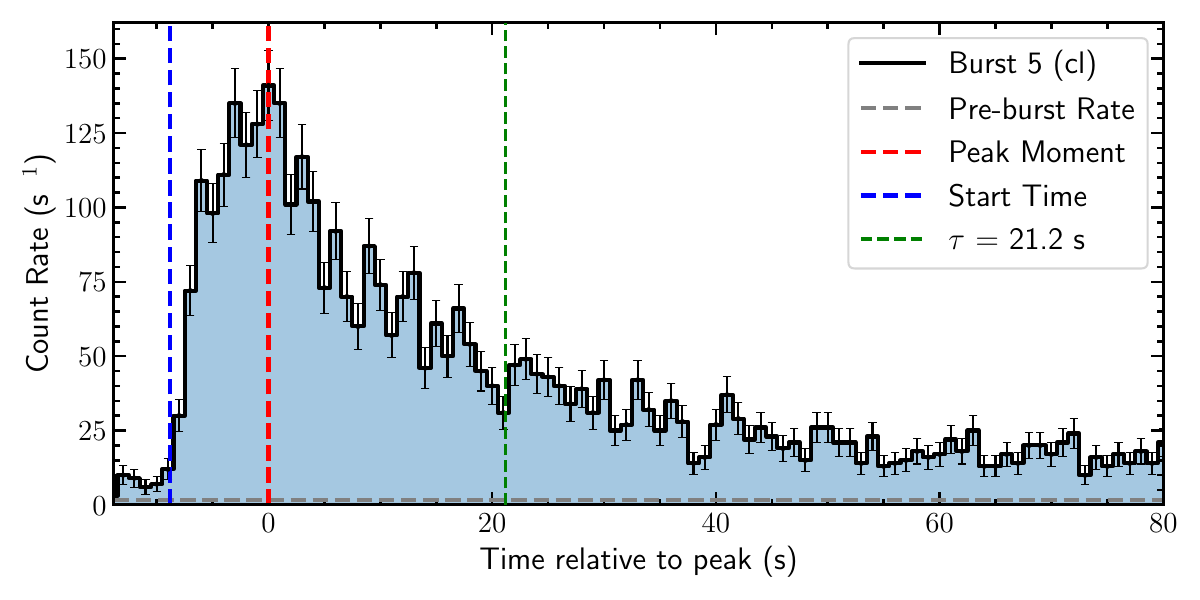}\\
\includegraphics[scale=0.4]{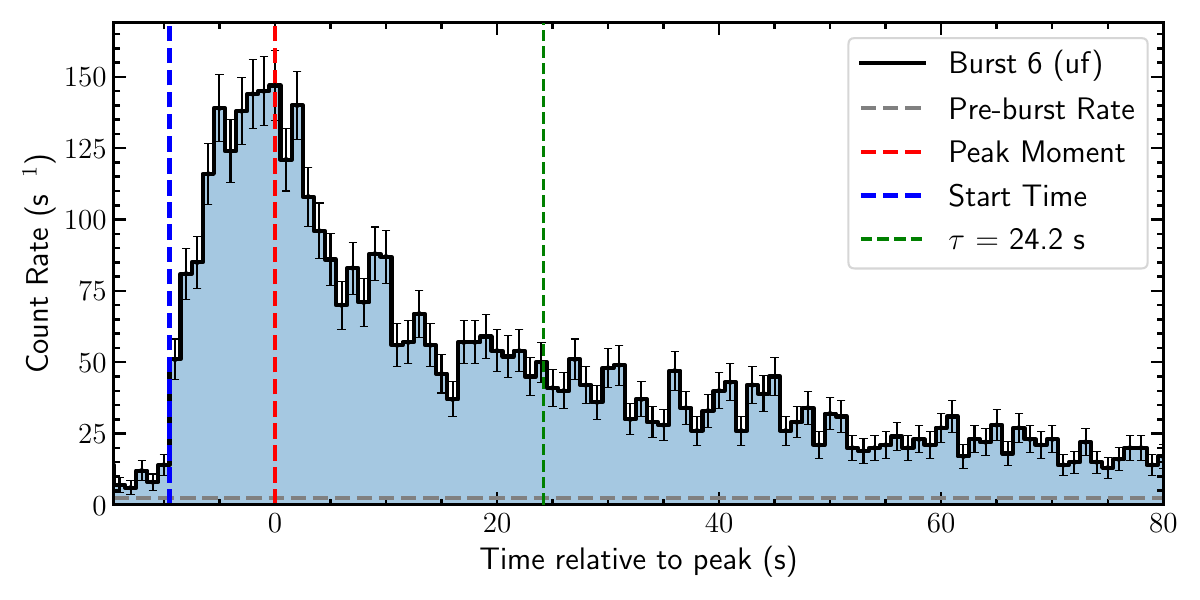} 
\includegraphics[scale=0.4]{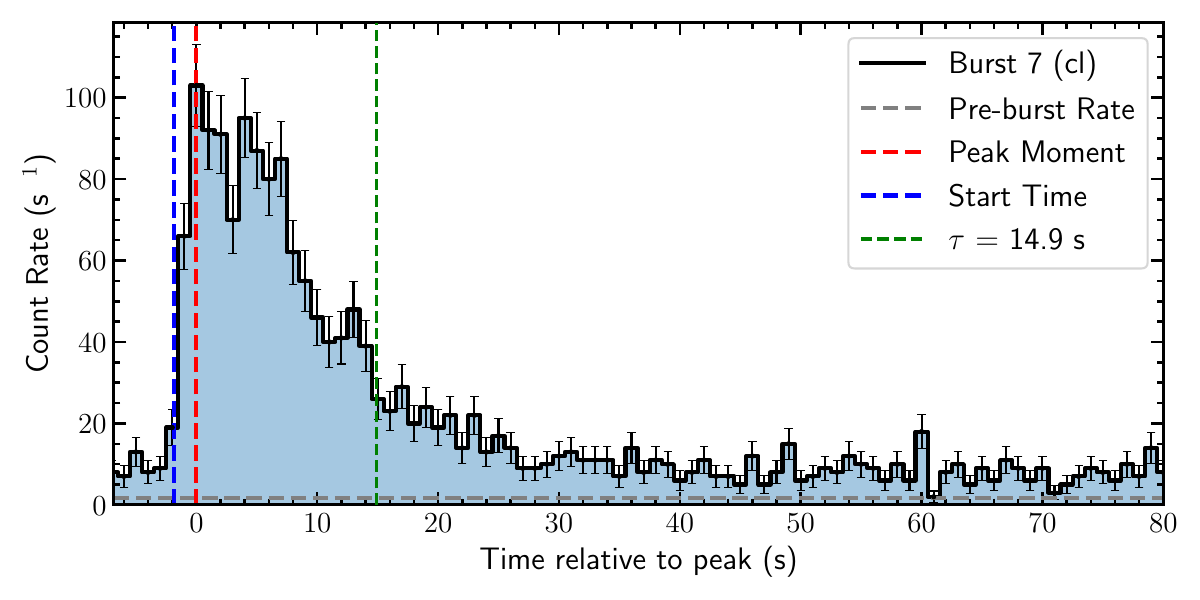} \\
\includegraphics[scale=0.4]{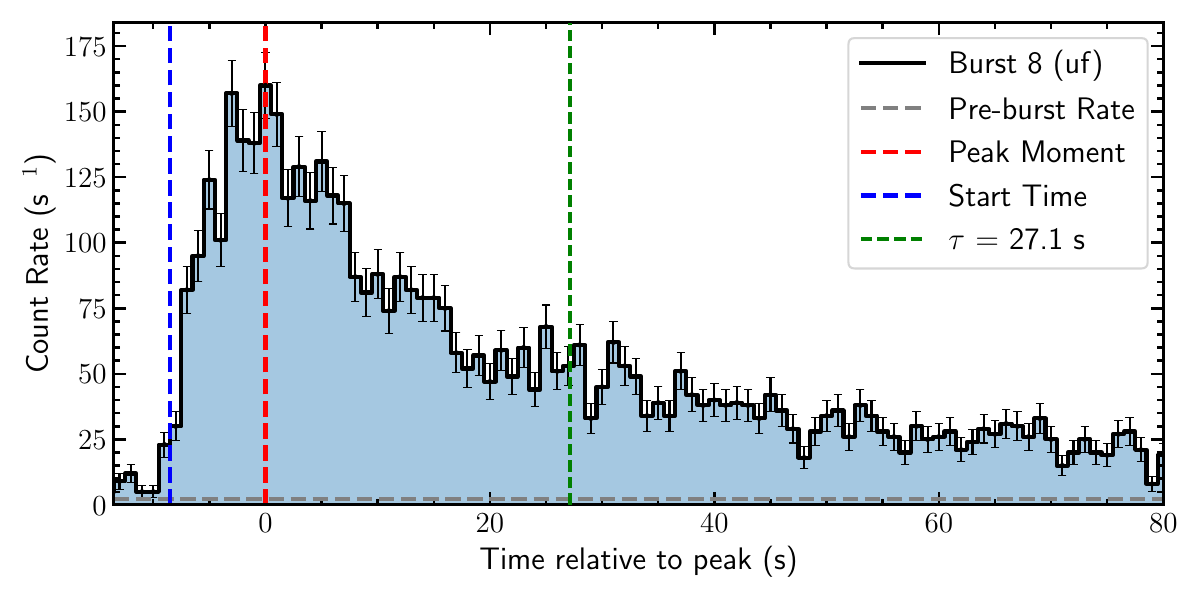} 
\includegraphics[scale=0.4]{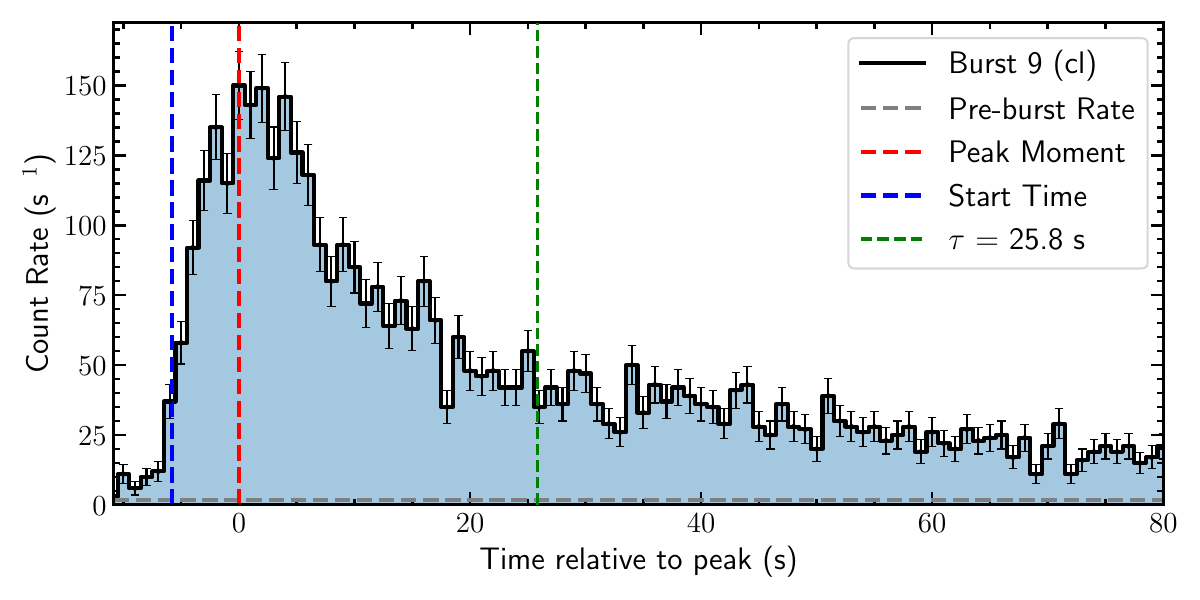}\\ 
\includegraphics[scale=0.4]{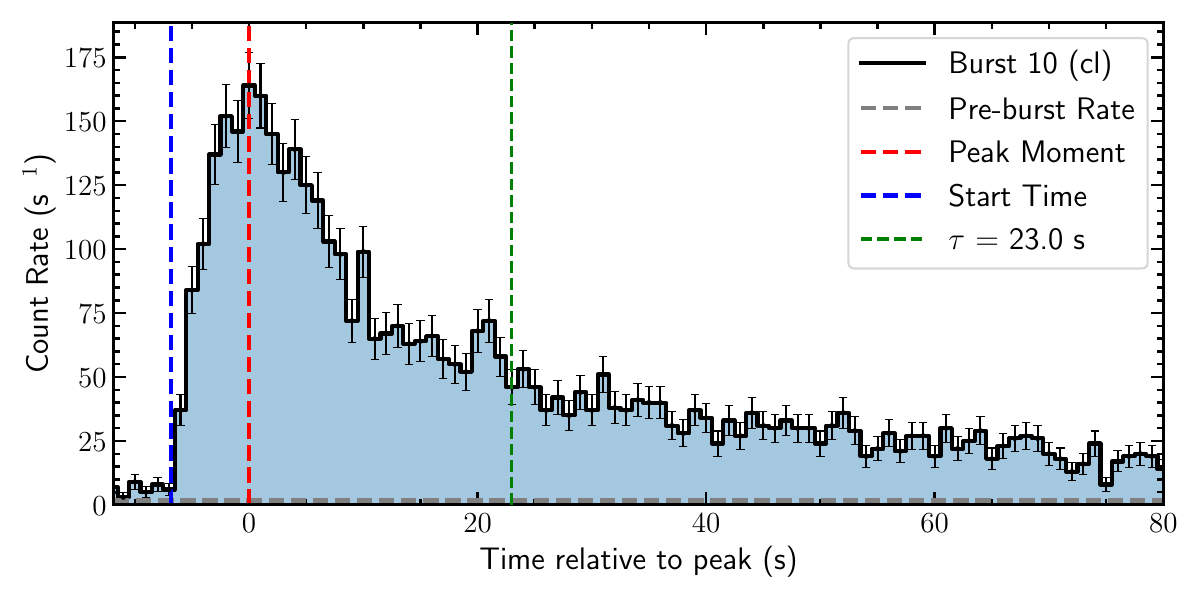} 
\caption{3.0--30.0~keV lightcurves of the detected thermonuclear X-ray bursts by \nustar. Bursts \#6 and \#8 are only seen in the unfiltered event files.}
\label{fig:nustarburst_lc}
\end{figure*}

\section{Spectral Evolution of the X-ray Bursts}

Time resolved spectroscopic evolution of the detected thermonuclear bursts with \xmm are shown. For bursts \#5, \#7, \#9 and \#10 our results show the simultaneous fits in the 1--20~keV range including the \nustar data as well, while for the others we only used 0.5--10.0~keV data from \xmm.

\begin{figure*}
\centering
\includegraphics[scale=0.53]{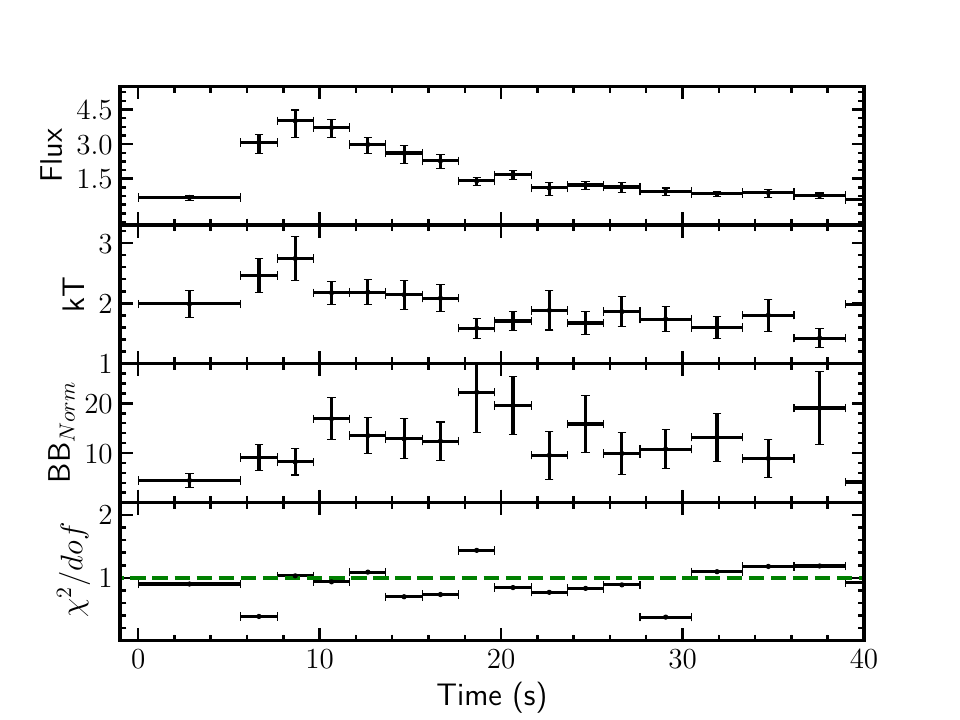}
\includegraphics[scale=0.53]{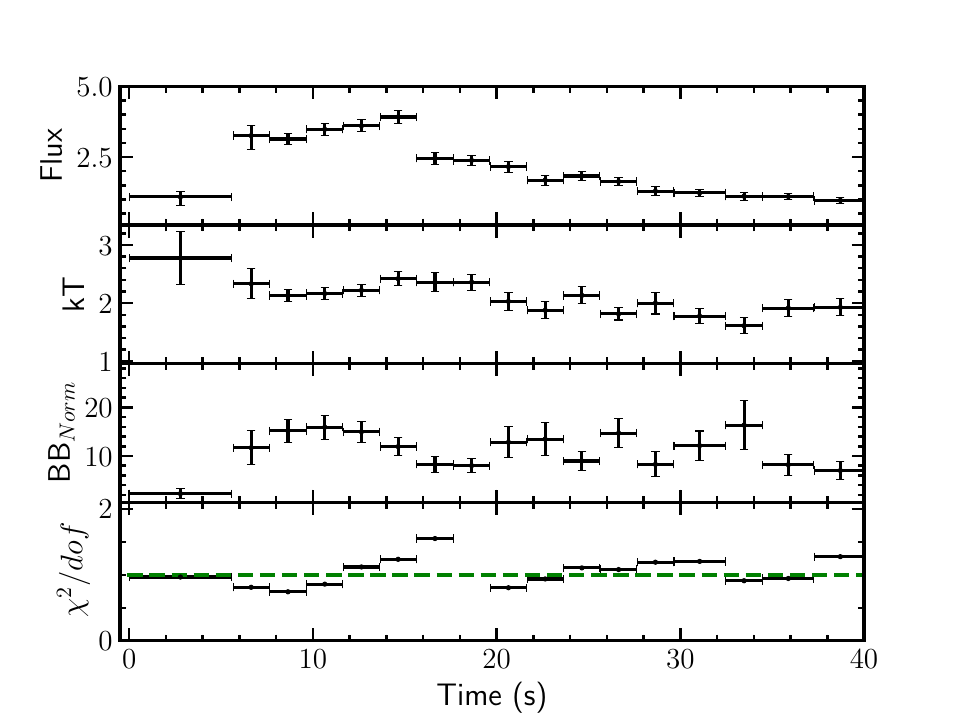}
\includegraphics[scale=0.53]{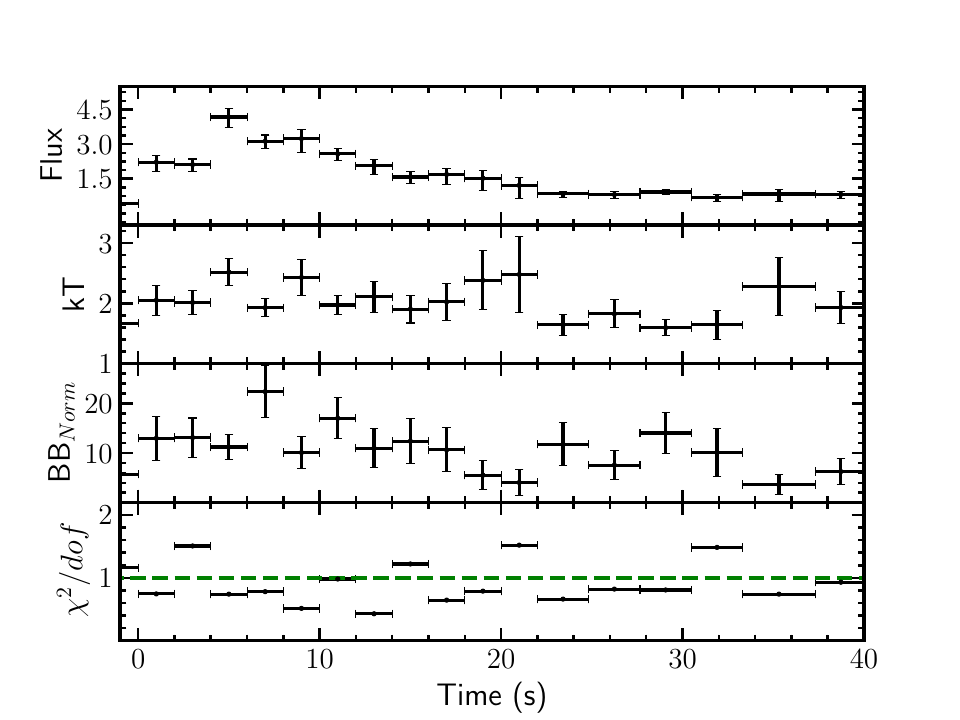}
\includegraphics[scale=0.53]{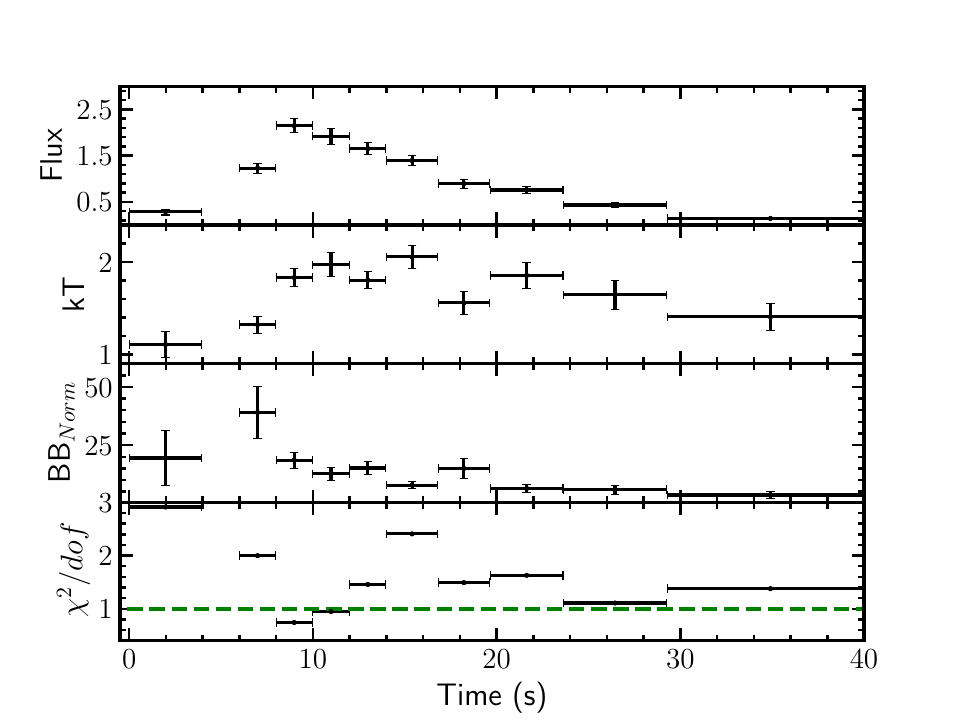}
\includegraphics[scale=0.53]{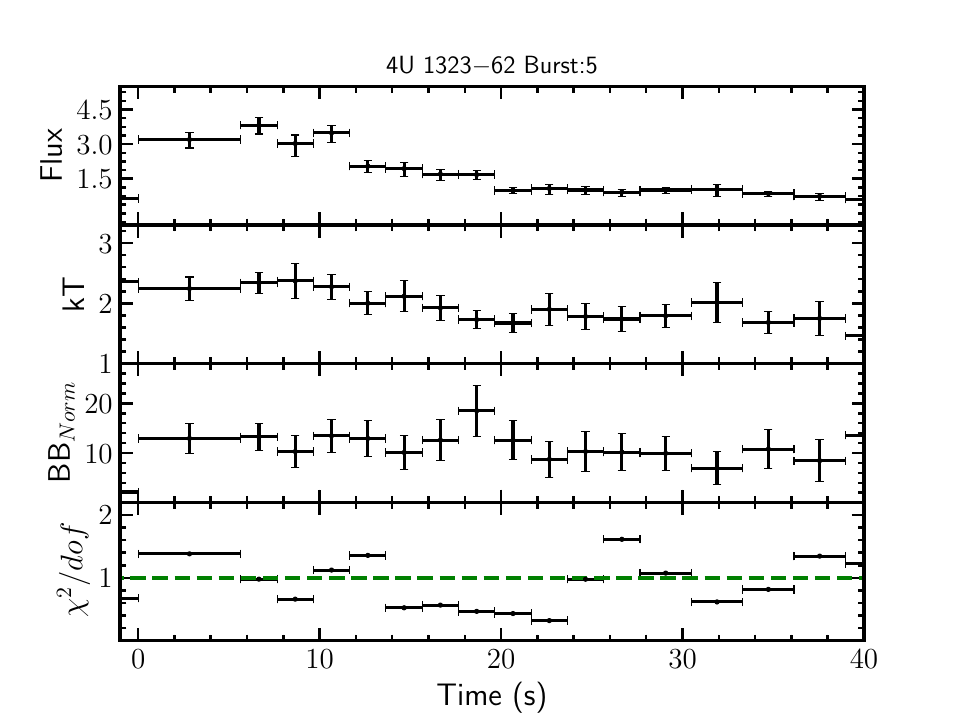}
\includegraphics[scale=0.53]{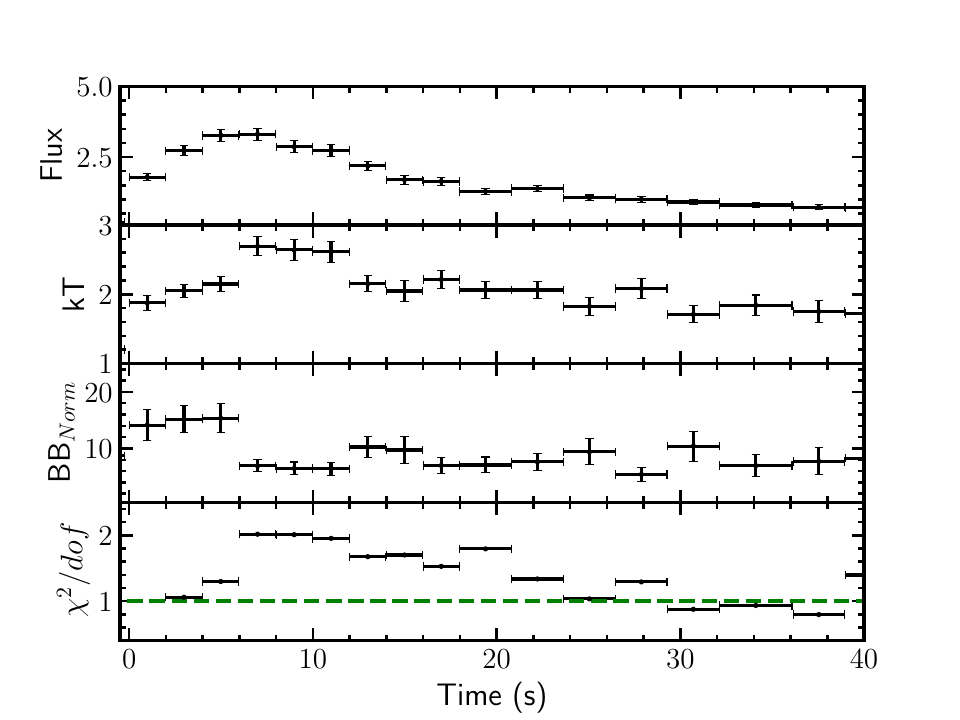}
\caption{Spectral evolution of the bursts (\#4 to \#9) from \src observed with \xmm. For each burst, from upper to lower panels, total bolometric unabsorbed flux (in units of \fluxcgs), blackbody temperature (in units of kT), normalization of the blackbody (in units of R$^2/D_{10 kpc}^2$) and the reduced $\chi^2$ is given respectively.}
\label{fig:ts_res}
\end{figure*}
\begin{figure*}
\centering
\includegraphics[scale=0.53]{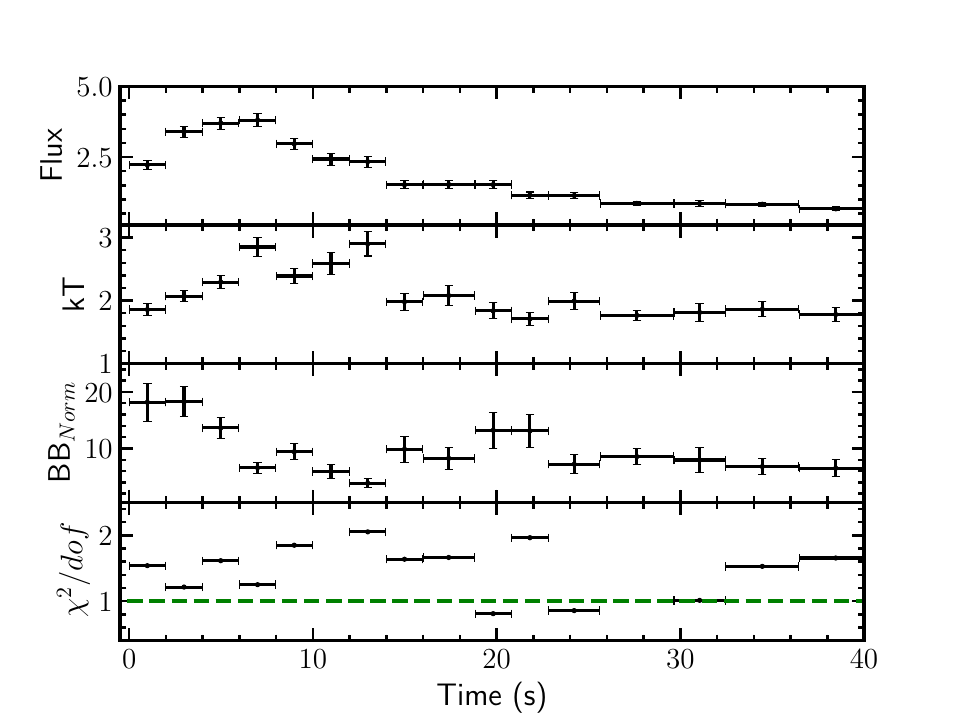}
 \includegraphics[scale=0.53]{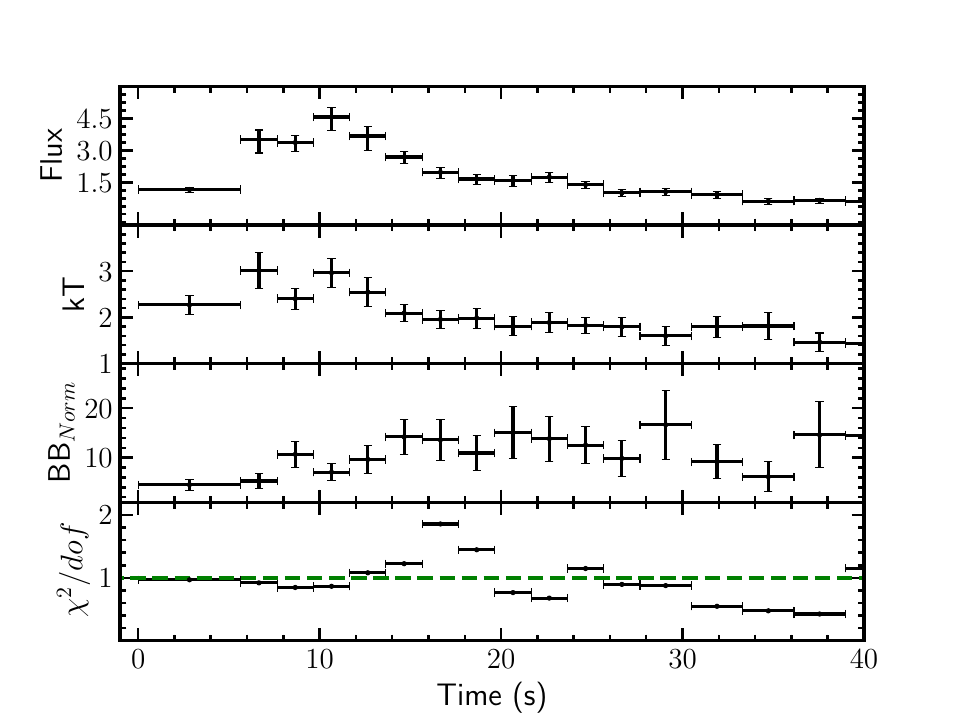}
\includegraphics[scale=0.53]{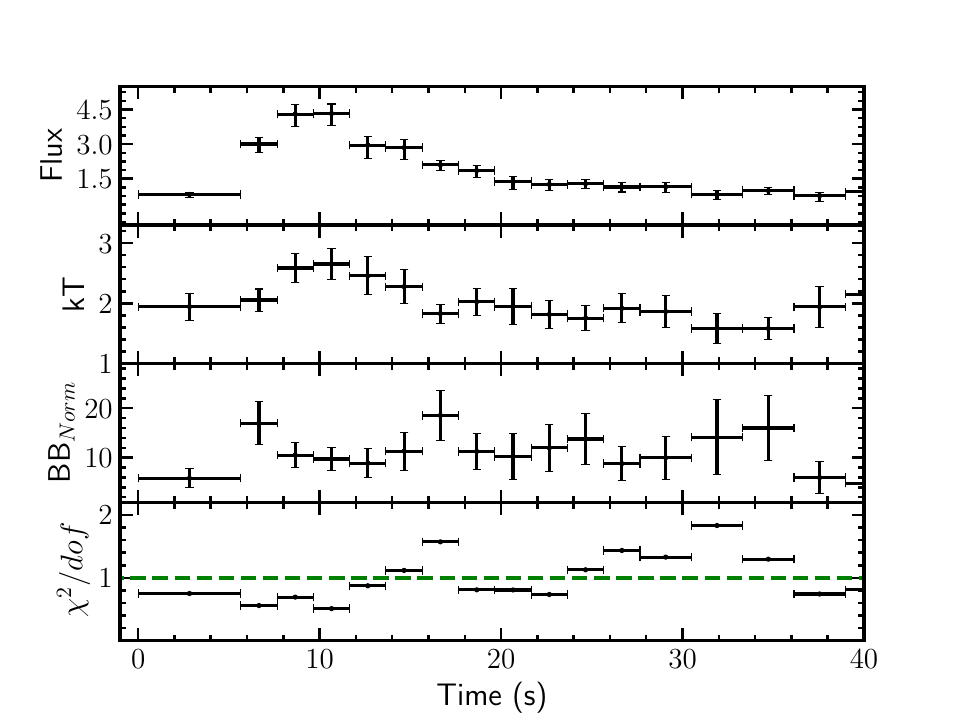}
\caption{Same as \autoref{fig:ts_res} but for bursts \#10, \#11, and \#12.}
\label{fig:ts_res2}
\end{figure*}

\section{X-ray Dips}
In this section we show the zoomed in views of the detected X-ray dips with \xmm from \src.

\begin{figure*}
    \centering
    \includegraphics[scale=0.5]{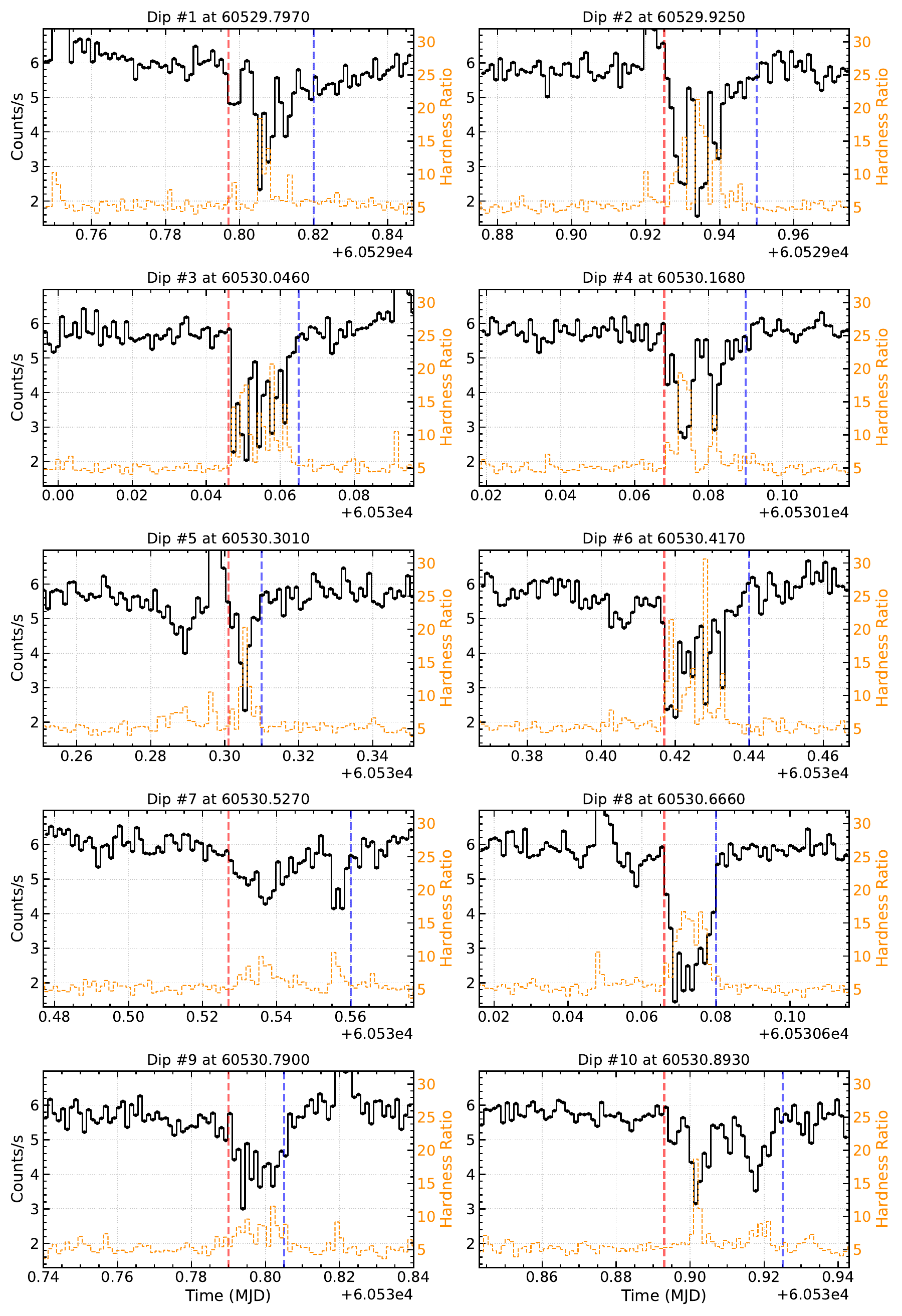}
    \caption{Zoomed in view of all the X-ray dips observed with \xmm in the 0.6--10.0~keV band from \src. Hardness ratios are also overplotted in orange. Start and end times are marked with vertical red dashed lines.}
    \label{fig:dipszoom}
\end{figure*}



\bsp	
\label{lastpage}
\end{document}